\documentclass[12pt,a4paper]{article}

\RequirePackage[l2tabu, orthodox]{nag}

\usepackage{mjheppub}

\usepackage[utf8]{inputenc}

\usepackage{amsmath,amsfonts,amssymb,epsfig,fp,youngtab,color,wasysym,bbm,hyperref}

\usepackage{tikz}
\usetikzlibrary{calc}
\usepackage{pgfplots}
\usepgfplotslibrary{fillbetween}
\usepackage{braids}

\newcommand{\nc}{\newcommand}
\nc{\rnc}{\renewcommand}
\nc{\nn}{\nonumber}
\nc{\del}{{\partial}}
\rnc{\Im}{{\rm{Im}\,}}
\rnc{\Re}{{\rm{Re}\,}}
\nc{\db}{\displaybreak[0]\\}
\nc{\bra}{\langle}
\nc{\ket}{\rangle}

\nc{\lam}{\lambda}
\nc{\g}{{\mathfrak{g}}}
\nc{\zb}{\bar{z}}
\nc{\hb}{\bar{h}}
\nc{\J}{\mathcal{J}}
\nc{\su}{\widehat{\mathfrak{su}}(2)_k}

\nc{\tcr}{\textcolor{red}}

\def\nn{\nonumber}
\def\beq{\begin{equation}}
\def\eeq{\end{equation}}
\def\bea{\begin{eqnarray}}
\def\eea{\end{eqnarray}}

\numberwithin{equation}{section}
\numberwithin{lemma}{section}
\numberwithin{proposition}{section}
\numberwithin{theorem}{section}
\numberwithin{corollary}{section}
\numberwithin{conjecture}{section}

\begin{document}

\title{Spin interfaces and crossing probabilities of spin clusters in parafermionic models}

\author{Yoshiki Fukusumi$^1$,}
\author{Marco Picco$^2$,}
\author{Raoul Santachiara$^3$}

\affiliation{$^1$ Department of Physics, Faculty of Science, University of Zagreb, HR-10000 Zagreb, Croatia}

\affiliation{$^2$ Sorbonne Universit\'e, CNRS UMR 7589, Laboratoire de Physique Th\'eorique et Hautes Energies, 
4 Place Jussieu, 75252 Paris Cedex 05, France}

%
\affiliation{$^3$ Université Paris-Saclay, CNRS, LPTMS, 91405, Orsay, France}


\abstract{
We consider fractal curves in two-dimensional $Z_N$ spin lattice models. These are $N$ states spin models that undergo a continuous ferromagnetic-paramagnetic phase transition described by the $Z_N$ parafermionic field theory. The main motivation here is to investigate the correspondence between Schramm-Loewner evolutions (SLE) and conformal field theories with extended conformal algebras (ECFT).
By using Monte-Carlo simulation, we compute the fractal dimension of different spin interfaces for the $N=3$ and $N=4$ spin models that correspond respectively to the $Q=3$ Potts model and to the Ashkin-Teller model at the Fateev-Zamolodchikov point. These numerical measures, that  improve and complete the ones presented in the previous works \cite{PSS,PS2}, are shown to be consistent with SLE/ECFT predictions. We consider then the crossing probability of spin clusters in a rectangular domain. Using a multiple SLE approach, we provide crossing probability formulas for $Z_N$ parafarmionic theories. The parafermionic conformal blocks that enter the crossing probability formula are computed by solving a Knhiznik-Zamolodchikov system of rank 3. In the $Q=3$ Potts model case ($N=3$), where the parafermionic blocks coincide with the Virasoro ones, we rederive the crossing formula found in  \cite{Flores_2017} that is in good agreement with our measures. For $N\geq 4$ where the crossing probability satisfies a third order differential equation instead of a second order one, our formulas are new. The theoretical predictions are compared to Monte-Carlo measures taken at $N=4$ and a fair agreement is found. 
}

\maketitle

\section{Introduction}

The study of the fractal objects appearing in critical phenomena, and the role of conformal invariance in describing their behavior, represents one of the most significant problems of current statistical and mathematical physics. Important insights into the nature of conformal fractals have been provided recently by studying the geometric \cite{prs19,saleur2020} and topological \cite{jps19two,javerzat2019fourpoint,javerzat2020topological} properties of percolation clusters. These results have been obtained using a conformal bootstrap approach that provides a new and powerful angle of attack to these problems. Here we consider different fractal objects: instead of fractal domains, we study  fractal curves defined as the boundaries of spin clusters. A very well known  method
to construct conformal fractal curves is the Schramm-Loewner Evolution (SLE)\cite{Schramm2000}. The SLE  correspondence to Conformal Field Theory (CFT) enables us to analyze 2D statistical models in a wide variety of geometries \cite{BAUER2002135} and should shed a unified understanding of boundary critical phenomena and stochastic processes. One of the main motivation behind this work is to investigate the  correspondence between the SLE and CFT with extended infinite symmetries (ECFTs) that adds to the conformal one \cite{wiegmann2005,SAKAI2013429}. An important example of ECFTs are the $Z_N$ parafermionic theories that describe the critical point of $Z_N$ invariant lattice spin models \cite{fzpara}. These models include the Ising ($N=2$), the  $3$-states spin Potts model ($N=3$) and the Ashkin-Teller model at the Fateev-Zamolodchikov point ($N=4$). The latter two models, defined in (\ref{H3potts}) and in (\ref{HATIs}) are the object of the Monte-Carlo simulations presented here. 

\noindent In \cite{Santachiara_2008,pisa07} the SLE/ECFT correspondence was used to predict the fractal dimension for spin interfaces in $Z_N$ spin models. The SLE approach to the entire critical line of the Ashkin-Teller model, which includes the $Z_4$ spin model, has been further discussed in \cite{Caselle_2011,Ikhlef_2010}. In \cite{fukusumi2017multiple}  multiple SLE processes in these theories were also considered. By Monte-Carlo simulations, we measure first the fractal dimension of different spin interfaces, 
showing that certain spin interfaces have a fractal dimension that is consistent with the predicted values. Note that these measures improve and complete the ones presented in previous papers \cite{PSS,PS2}. 

\noindent We study then the crossing probability of $Z_N$ spin clusters by using the multiple SLE approach of \cite{fukusumi2017multiple}. The analytical predictions are then compared   to  Monte-Carlo measures, taken for the $Z_3$ and $Z_4$ spin lattice model on rectangular domain. The case $N\geq 4$ is particularly interesting to further test the SLE/ECFT correspondence. Indeed, the role of the extended symmetry becomes important and the formulas deviate from the ones concerning CFTs based on Virasoro algebras only. Moreover, at the level of the $Z_N$ spin lattice models, for $N\geq 4$ the identification of all the boundary conformal states in terms of spin configuration is not known. In the simulations, we test different boundary conditions that generate a single spin interfaces and we select the ones that are better described by the fractal dimension predicted by the SLE/ECFT correspondence.   

In section $2$, we define the models and the spin interfaces we simulate. We 
present new numerical  results for the fractal dimension of spin interfaces for the $Q=3$ Potts model
and the $Z_4$ spin model.   
In section $3$ we consider the crossing probability of $Z_N$ spin cluster in rectangular domains.
We compare analytical prediction to numerical data for the $Q=3$ Potts model
and the $Z_4$ spin model. The details of the computation of the parafermionic blocks entering the $Z_N$ crossing formula are given in the Appendixes. We summarise the results in the Conclusion.

\section{The $Z_N$ spin interfaces and their fractal dimension}
\label{disc}

In this paper, we will present accurate measurements of fractal dimensions for spin interfaces and crossing probabilities 
for the $Z_3$ and  $Z_4$ parafermionic spin models on square lattice of rectangular shape. 

\subsection{The $Z_3$ and $Z_4$ spin models}
The $Z_3$ parafermionic spin model, that coincides with the $Q=3$ spin Potts model, is defined by the Hamiltonian:
\begin{equation}
H_{Z_3} =-\sum_{<ij>} K \delta_{S_{i}S_{j}},
\label{H3potts}
\end{equation} 
where $S_{i}$ is the spin variable taking values $S_{i}=1,2,3$ and the sum is restricted to the neighbouring sites $<ij>$. The ferromagnetic-paramagnetic transition is located, for the square lattice, at $K_c= \log(1 +\sqrt{3})$ and it is described by the $Z_3$ parafermionic theory that has central charge $c=4/5$.  
The thermal and magnetic sector of this theory coincides with the one of the Virasoro (non-diagonal) minimal model of the $D-$ series \cite{prs19}.  

The $Z_4$ parafermionic spin model coincides with the Ashkin-Teller (AT) model at a particular values of the couplings. This latter   model \cite{AT43} can be 
defined in terms of two coupled Ising models: on each site $i$ of a square lattice one
associates a pair of spins, denoted by $\sigma_i$ and $\tau_i$, which
take two values, say up(+) and down(-). The Hamiltonian is defined by
\begin{equation}
H_{Z_4} =-\sum_{<ij>} K (  \sigma_i \sigma_j + \tau_i \tau_j )  + K_4 \sigma_i \sigma_j \tau_i \tau_j \; .
\label{HATIs}
\end{equation}
The two parameters, $K$ and $K_4$, correspond
respectively to the usual Ising spin interaction and to the 4-spins
coupling between two Ising models. Using the map:
\bea
\label{def:mapSs}
S_i=1 \rightarrow \sigma_i = + , \tau_i = + \; \; ; \; \;  S_i=2 \rightarrow \sigma_i = + , \tau_i = - \nn \\ \nn
S_i=3 \rightarrow \sigma_i = - , \tau_i = - \; \; ; \; \;  S_i=4 \rightarrow \sigma_i = - , \tau_i = + \; ;  
\eea
one can rewrite the AT model in terms of a one layer of spins $S_i$ that take four values and interact via next-neighbours interactions.  
The AT model, that on a square lattice  is equivalent to the staggered six vertex model~\cite{N1984,SaleurAT}, presents a rich phase diagram~\cite{Baxter,N1984,SaleurAT}. There exists a critical line which is defined  by
the self-dual condition 
\begin{equation}
\sinh 2 K = \exp (-2 K_4)
\end{equation}
and terminates at $\coth 2 K_2=2$.  The point on the critical line defined by:
\beq
 x_1^{\mbox{FZ}}= {\sin({\pi \over 16}) \over \sin({3 \pi\over 16} )} \; 
; \; x_2^{\mbox{FZ}}= x_1 {\sin({5 \pi \over 16}) \over \sin({7\pi \over 16})} \; ,
\label{FZ} 
\eeq
where $x_1$ and $x_2$ are related to $K$ and $K_4$~:
\begin{equation}
\exp (4 K)={1+2 x_1 + x_2 \over 1 - 2 x_1 + x_2} \quad ; \quad \exp (2 K+ 2 K_4)=\frac{1+2 x_1 + x_2 }{1-x_2} \; .
\end{equation}
is called the Fateev-Zamolodchikov point. At this point the model has been shown to admit special integrable properties
\cite{FZ,Yang-Perk} and to be described, in the continuum limit, by the $Z_4$ parafermionic theory.

One can define different types of interfaces in these two models according to the colors of spins they separate. For instance, in the $Q=3$ Potts model, the interface $(1|23)$ separates the spin of color $1$ from the spins of colors $2$ or $3$. Or, in the $Z_4$ spin model, the interface $(12|34)$ separates the spin of color $1$ or $2$ to the spins of color $3$ or $4$. These interfaces are defined in the bulk. The same notation is used for the spin boundary conditions that generate a single boundary interface. We consider the following boundary conditions. For the $Q=3$ Potts model:
\begin{itemize}
\item $(1|23)$: this corresponds to fix $S_i=1$ on the  top and bottom boundaries 
(of length $L_x$)  while the spins on the two other boundaries can  take all values different from $1$
\end{itemize}
Note that, in the case of Virasoro minimal models, crossing probability formulas for general polygonal domains and general central charges,  have been given in \cite{Flores_2017}. These formulas include as a special case  the formula  (\ref{Q=3CP}) derived here. For the $Z_4$ model, natural choices of boundary conditions are: 
\begin{itemize}
\item $(1 | 234)$: from (\ref{def:mapSs}), this corresponds to fix $S_i=1$ on the top and bottom boundaries 
(of length $L_x$)  while the spins on the two other boundaries can  take all values different from $1$, i.e.  $S_i \in (2,3,4)$. In terms of the Ising degrees of freedom, this is equivalent  to set 
$\sigma_i = + $ and $\tau_i = + $ on the top and bottom boundaries and any other choice 
on the two other boundaries. 

\item $(12 | 34)$ : on the top and bottom boundaries, $S_i \in (1,2)$  on the left and right boundaries $S_i \in (3,4)$. This corresponds to impose $\sigma_i = + $ on the top and bottom boundaries while $\sigma_i = - $ on the other two boundaries. The same conditions have been considered in \cite{Caselle_2011} for the entire critical line of the AT model.

\item $(13|24)$: on the top and bottom boundaries $S_i \in (1,3)$, on the left and right boundaries $S_i \in (2,4)$. This corresponds  to $\sigma_i = +, \tau_i = +$ or $\sigma_i = -, \tau_i = -$ on the top and bottom boundaries 
and $\sigma_i = +, \tau_i = -$ or $\sigma_i = -, \tau_i = +$ on the two other boundaries. 

\end{itemize}
Due to the symmetry of the system, it is easy to see that these are the only conditions that can be considered. There exists 
also conditions like $(1 | 2)$ but then there can be no connecting cluster between opposite boundaries. We will not consider these cases here.

\subsection{Spin interfaces in the $Q=3$ Potts model}

The boundary conditions  ($1|23$) were considered in~\cite{GC}. For this model, the authors predicted the fractal dimension of the interface to be $d_f =1 + (10/3)/8 $ and indeed found this result when considering fluctuating boundary conditions 
with a single interface. They also consider the case with fixed boundary conditions $(1|2)$.  
On one part of the boundary, the spins are  fixed to the type $S_i=1$ while the spins on the remaining 
boundary are fixed to the value $S_i=2$. The generated interface can then be separated in two parts~: 
a first one, of length $l_{\rm composite}$, separates spins of color $S=1$ connected to the boundary with $S=1$ and 
spins of color $S=2$ connected to the boundary $S=2$ and is called the composite part; 
the remaining part of the interface, of length $l_{\rm split}$, which touch spins of color $S=3$ or isolated clusters is called the split part. The fractal dimensions are not the same for the two parts. Gamsa and Cardy obtained $d_f \simeq 1.0$ for the composite part and $d_f \simeq 1.6$ for the split part ~\cite{GC}.
More accurate results \cite{PS4} show that in fact this is not the case. By considering larger sizes and more statistics, 
it can be seen that the split interface is composed of two parts.  

\begin{figure}
\begin{center}
\epsfxsize=350pt\epsfysize=300pt{\epsffile{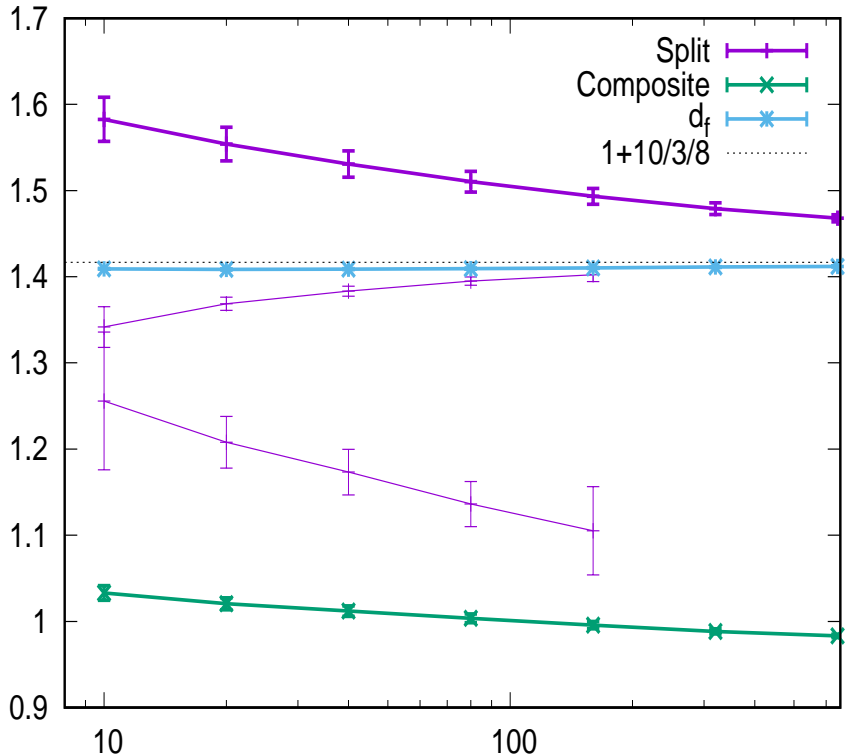}}
\end{center}
\caption{
Effective fractal dimensions for the $Q=3$ state Potts model as a function of the smallest size $L$ in the fit. See the text for details.
\label{F3}
}
\end{figure}

In Fig.~\ref{F3} we show effective fractal dimensions as a function of $L$ in a fit with data between the size $L$ and the maximum size $L_m=5120$. In this figure, we show various quantities. First the fractal dimension for the total interface 
 $l_{\rm composite} + l_{\rm split} \simeq L^{d_f}$. It converges very nicely towards the expected value 
 $d_f = 1+(10/3)/8$. Next, the fractal dimension for the split interface which has an effective exponent decreasing very slowly. 
 At intermediate sizes, the value is close to the one obtained by Gamsa and Cardy but for the largest sizes that we simulate, it goes to smaller values. We also show that the fractal dimension for the composite interface which decreases slowly down to $\simeq 0.98$. Again, at intermediate sizes, it is compatible with the finding of Gamsa and Cardy. 
Next, for the split interface, we show the result of a fit with two exponents as in eq.~(\ref{fractd2}). 
These two effective exponents are shown with thin lines with the same color. 
We observe that one of the exponents goes towards the fractal dimension of the fluctuating boundary conditions $d_f = 1+(10/3)/8$. The second fractal dimension seems to converge towards the value of the fractal dimension of the composite interface. This last measurement is plagued by large error bars, 
as can be expected since it is a subdominant exponent. 

Then, for the $Q=3$ Potts model, the split interface is just the trivial sum (or more precisely difference) of two scaling interfaces. 

\subsection{Spin interfaces in the $Z_4$  model}
\label{z4inter}
We will show in the following that analogous results are obtained for the $Z_4$ model.

In \cite{PSS,PS2,PS3}, the fractal dimension of the interfaces based on boundary conditions 
$(1|234)$, $(12|34)$ and $(13|24)$ were considered. In these numerical measurements, even 
for the largest systems considered, we observed important finite size corrections. Then it was not possible 
to determine with a great precision the fractal dimension. In these simulations, the main limitation was due to the 
fact that we considered lattices with fixed boundary conditions. This has the consequence to slow down the Monte Carlo updates 
and then limit the precision one can get. We present here some alternative way for performing similar measurements with some new results. 

Another method is to determine the fractal dimensions of bulk interfaces. This was already done in \cite{PSS} where we considered the fractal dimensions of finite clusters. In that case, one needs to compare the average volume of the clusters with the average length of the interface surrounding them, which makes the measurement indirect. In this paper, we propose a different way of doing this measurement. We will simulate systems with periodic boundary conditions and for each independent configurations, we will consider the largest cluster. More precisely, we will compute the largest cluster ${\cal C}_1$ containing only one value of the spin $S_i$, the largest cluster ${\cal C}_{12}$ containing only states with $S_i=1$ or $2$ and the largest cluster ${\cal C}_{13}$ containing only states with $S_i=1$ or $3$. 
Since we are at a critical point, the probability that a cluster of any type percolates is finite. If one considers a lattice with periodic boundary conditions, then there is a finite probability that this percolating cluster is wrapping through the lattice such that the borders of the cluster correspond to two interfaces. In this new approach, we measure the average length of these interfaces and we use them to define the fractal dimension via the relation 
\beq
\label{fractd}
l \simeq L^{d_f}
\eeq
where $l$ the length of the interface and $L$ the lattice size. 
\begin{figure}
\begin{center}
\epsfxsize=300pt\epsfysize=240pt{\epsffile{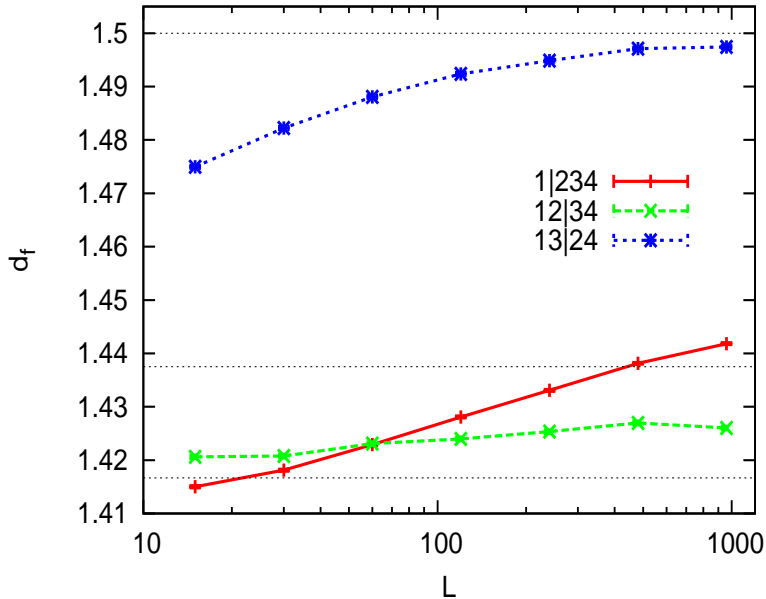}}
\end{center}
\caption{
Fractal dimension $d_f$ as a function of ${L}$ for the conditions $(1|234)$, $(12|34)$ and $(13|24)$ for the ${\cal  Z}_4$ parafermion model. The dashed lines correspond to $1+10/24$, $1+7/16$ and $3/2$.  
\label{F2}
}
\end{figure}
The advantage of this method is that the auto-correlation time is much smaller than in 
the case with fixed boundary conditions. For example, for $L=640$, the auto-correlation  time is determined to be $\tau=356$ while it is $\tau \simeq 25000$ for the case with an interface generate by fixed boundary conditions $(1|234)$. 
This allows us to better compute $d_f$ for the various cases, achieving a much better accuracy. 
In Fig.~\ref{F2}, we show the effective fractal dimension for the ${\cal  Z}_4$ parafermion model obtained for the various cases using the following formula~:
\beq
d_f^{eff} (1.5 L ) = { \log{(l(2 L)/ l(L))} \over \log{2}} \; .
\eeq
Notice that in Figure~\ref{F2}, error bars are very small.
\begin{figure}
\begin{center}
\epsfxsize=300pt\epsfysize=240pt{\epsffile{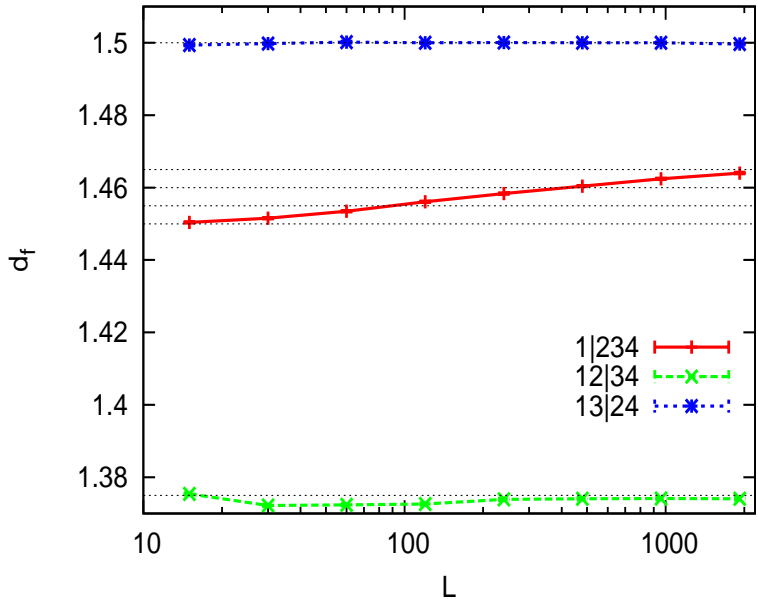}}
\end{center}
\caption{
Fractal dimension $d_f$ as a function of ${L}$ for the conditions $(1|234)$, $(12|34)$ and $(13|24)$ for two decoupled Ising models (also known as XOR model).  
The dashed lines correspond to $1+3/8$, $3/2$ on the right. We also add some dashed lines at intermediate values as a guide to the eye. 
\label{F2Is}
}
\end{figure}
This was made possible by accumulating a lot of data, namely $10^8$ independent configurations. The total computing time for 
the largest run ($L=1280$ of the ${\cal  Z}_4$ parafermion model) corresponds to 100 years of run on a single processor.

For the condition $(13|24)$ we observe a nice convergence towards the value $3/2$, consistently with was already obtained in previous works \cite{PS2,PS3}. For the condition $(12|34)$, we observe that, even if  the finite size corrections are still visible, it seems reasonable to expect that the large size limit will be in the range $1.42-1.44$. This is compatible with the prediction $1+7/16$, discussed previously. 
For the condition $(1|234)$ we observe that the effective fractal dimension is constantly increasing with no apparent saturation. It is difficult to predict the asymptotic 
limit, but one can expect that it will saturate towards $1.5$. Similar measurements were also done for the decoupled point of the Ashkin-Teller line and a similar behavior is  observed, see Fig.~\ref{F2Is}. For that point, the fractal dimension for the condition $(1|234)$ does also not converge and moves slowly towards 
the value $3/2$. The fractal dimensions for the two other conditions converge nicely to the expected dimensions, respectively to $d_f = 1 + 3/8$ and $d_f =3/2$. 

A simple scenario to interpret the above numerical observation is that the condition $(1|234)$ corresponds to a mixing of two conformal boundary conditions. The dominant one would have a dimension $3/4$ and the second  a smaller dimension with a value close to the condition $(12|34)$.  As a check, we replaced the relation (\ref{fractd}) by 
\beq
\label{fractd2}
l \simeq L^{d_{f_1}} + L^{d_{f_2}} \; .
\eeq
In the case of two decoupled Ising model, we obtain a good fit of the length of the interface $l$ for the boundary condition 
$(1|234)$ with the fixed value $d_{f_2}=1.5$ and with $d_{f_1} \simeq 1.38$, thus a value close to the exact value for $(12|34)$. 
Equivalently, we can fix $d_{f_1}=1.375$. A fit to the form (\ref{fractd2}) gives then $d_{f_2} \simeq 1.49$ thus close to the value for $(13|24)$. 
Fitting both fractal dimensions gives comparable results but with very large error bars. 

We also checked a similar type of fit for the data corresponding to the ${\cal  Z}_4$ parafermion model). We obtain then 
$d_{f_2} \simeq 1.46-1.52$ and $d_{f_1} \simeq 1.25-1.35$. For this model, the errors are much larger than for the two decoupled Ising model, so we can just claim that this is compatible with our mixing scenario.

\section{$Z_N$ crossing probability}
We present now results for the measurement of crossing probabilities in the case of the $Z_3$ and $Z_4$ spin models.
In both cases, we consider lattices with the geometry $L_h \times L_v$ with $L_h = r L_v$ and we impose some fixed conditions on the left and right boundaries and another fixed condition on the top and bottom boundaries. 
The four corners are left empty, since they are common to two borders with different boundary conditions. 
We compute the probability that a cluster connects the left and right boundaries or 
that a cluster connects the top and bottom boundaries. If the conditions on the boundaries are exclusive, 
then there must exist one only one connecting cluster, either from left to right $(P_h=1,P_v=0)$, or from bottom to top 
with, $(P_h=0, P_v=1)$. Taking into account some details concerning crossing in square lattice rectangular domains, discussed below in Section \ref{measurements}, we define the crossing probabilities as:
\beq
\label{defP}
{\cal P}_h= {\left< P_h\right> \over \left<P_h\right> + \left<P_v\right>} \; \; ; \; \; {\cal P}_v= {\left< P_v\right> \over \left<P_h\right> + \left<P_v\right>},
\eeq
where  $\left<\cdots \right>$ indicates the average over all the independent configurations.

\subsection{Theoretical predictions}

To compute crossing probabilities, we use the multiple-SLE approach discussed in \cite{fukusumi2017multiple}, which represents a generalization to the Virasoro case,  explained in detail in Section 8.2 of \cite{Bauer2005}.  We show first how to re-derive the result of \cite{Flores_2017} for the crossing probability for the $Q=3$ spin interface. Let us consider this model on the half plane. On the real line the alternating boundary conditions $(1|23)$  generate  single interfaces, as shown in the following Figure

\begin{center}
\begin{tikzpicture}
\draw[thick] (-3,0.5)--(13,0.5);
\draw[black,fill] (-2.5,0.5) circle [radius=0.1];
\draw[black,fill] (-1.5,0.5) circle [radius=0.1];
\draw[black,fill] (-0.5,0.5) circle [radius=0.1];
\draw[black,fill] (0.5,0.5) circle [radius=0.1];
\draw[blue,fill] (-2.5,1.5) circle [radius=0.1];
\draw[black,fill] (-1.5,1.5) circle [radius=0.1];
\draw[black,fill] (-0.5,1.5) circle [radius=0.1];
\draw[blue,fill] (0.5,1.5) circle [radius=0.1];
\draw[red,fill] (-2.5,2.5) circle [radius=0.1];
\draw[black,fill] (-1.5,2.5) circle [radius=0.1];
\draw[blue,fill] (-0.5,2.5) circle [radius=0.1];
\draw[blue,fill] (0.5,2.5) circle [radius=0.1];
\draw[black,fill] (-2.5,3.5) circle [radius=0.1];
\draw[black,fill] (-1.5,3.5) circle [radius=0.1];
\draw[blue,fill] (-0.5,3.5) circle [radius=0.1];
\draw[red,fill] (0.5,3.5) circle [radius=0.1];
\draw (1,0.25)--(1,1)--(0,1)--(0,2)--(-1,2)--(-1,3.75);
\draw(1,0.25) node[below]{0};
\draw(-1,-0.25) node[below]{$S_i=1$ };
\draw[blue,fill] (1.5,0.5) circle [radius=0.1];
\draw[red,fill] (2.5,0.5) circle [radius=0.1];
\draw[red,fill] (3.5,0.5) circle [radius=0.1];
\draw[blue,fill] (4.5,0.5) circle [radius=0.1];
\draw[black,fill] (1.5,1.5) circle [radius=0.1];
\draw[red,fill] (2.5,1.5) circle [radius=0.1];
\draw[black,fill] (3.5,1.5) circle [radius=0.1];
\draw[black,fill] (4.5,1.5) circle [radius=0.1];
\draw[blue,fill] (1.5,2.5) circle [radius=0.1];
\draw[blue,fill] (2.5,2.5) circle [radius=0.1];
\draw[black,fill] (3.5,2.5) circle [radius=0.1];
\draw[black,fill] (4.5,2.5) circle [radius=0.1];
\draw[blue,fill] (1.5,3.5) circle [radius=0.1];
\draw[black,fill] (2.5,3.5) circle [radius=0.1];
\draw[black,fill] (3.5,3.5) circle [radius=0.1];
\draw[black,fill] (4.5,3.5) circle [radius=0.1];
\draw (5,0.25)--(5,1)--(3,1)--(3,3)--(1,3)--(1,3.75);
\draw(5,0.25) node[below]{x};
\draw(3,-0.25) node[below]{$S_i=2$ or $3$};
\draw[black,fill] (5.5,0.5) circle [radius=0.1];
\draw[black,fill] (6.5,0.5) circle [radius=0.1];
\draw[black,fill] (7.5,0.5) circle [radius=0.1];
\draw[black,fill] (8.5,0.5) circle [radius=0.1];
\draw[red,fill] (5.5,1.5) circle [radius=0.1];
\draw[black,fill] (6.5,1.5) circle [radius=0.1];
\draw[black,fill] (7.5,1.5) circle [radius=0.1];
\draw[black,fill] (8.5,1.5) circle [radius=0.1];
\draw[blue,fill] (5.5,2.5) circle [radius=0.1];
\draw[black,fill] (6.5,2.5) circle [radius=0.1];
\draw[red,fill] (7.5,2.5) circle [radius=0.1];
\draw[blue,fill] (8.5,2.5) circle [radius=0.1];
\draw[black,fill] (5.5,3.5) circle [radius=0.1];
\draw[black,fill] (6.5,3.5) circle [radius=0.1];
\draw[red,fill] (7.5,3.5) circle [radius=0.1];
\draw[red,fill] (8.5,3.5) circle [radius=0.1];
\draw(7,-0.25) node[below]{$S_i=1$ };
\draw[red,fill] (9.5,0.5) circle [radius=0.1];
\draw[red,fill] (10.5,0.5) circle [radius=0.1];
\draw[red,fill] (11.5,0.5) circle [radius=0.1];
\draw[blue,fill] (12.5,0.5) circle [radius=0.1];
\draw[black,fill] (9.5,1.5) circle [radius=0.1];
\draw[red,fill] (10.5,1.5) circle [radius=0.1];
\draw[black,fill] (11.5,1.5) circle [radius=0.1];
\draw[black,fill] (12.5,1.5) circle [radius=0.1];
\draw[blue,fill] (9.5,2.5) circle [radius=0.1];
\draw[blue,fill] (10.5,2.5) circle [radius=0.1];
\draw[black,fill] (11.5,2.5) circle [radius=0.1];
\draw[black,fill] (12.5,2.5) circle [radius=0.1];
\draw[black,fill] (9.5,3.5) circle [radius=0.1];
\draw[red,fill] (10.5,3.5) circle [radius=0.1];
\draw[black,fill] (11.5,3.5) circle [radius=0.1];
\draw[black,fill] (12.5,3.5) circle [radius=0.1];
\draw (9,0.25)--(9,1)--(10,1)--(10,2)--(7,2)--(7,3.75);
\draw(9,0.25) node[below]{1};
\draw(11,-0.25) node[below]{$S_i=2$ or $3$};

\end{tikzpicture}
\end{center} 
The  $0,x$ and $1$ are the points where the boundary conditions change and where three simple ($1|23$) interfaces are generated. 
The partition function $Z$ can be seen as a sum of two partition functions:
\begin{equation}
Z(x)= Z_{I}(x)+Z_{II}(x),
\end{equation}  where $Z_{I}$ ($Z_{II}$) is the partition function of the configurations where the interface generating at $1$ (at $0$) collides with the point at infinity.
The boundary changing condition (b.c.c.) field associate to $(1|23)$ is the field $\varepsilon$ with dimension $\Delta_{\varepsilon}=\frac{2}{5}$. 
The $Z_{I}$ and $Z_{II}$ are related to the conformal blocks that contain  four $\varepsilon$ fields:
\begin{equation}
\label{CBpara3}
G_{3}(x)=\left< \varepsilon(0) \varepsilon (x)\varepsilon(1)\varepsilon(\infty)\right>_{Z_3}.
\end{equation}
\noindent 
The parafermionic block $G_{3}(x)$ coincides with the conformal block with four  Virasoro fields, with the same dimension, and degenerate at level two. We refer the reader to the Appendix of \cite{Santachiara_2008} for a detailed proof of this. In Appendix \ref{Ising} we show an analogous result for the Ising model ($N=2$).  $G_{3}(x)$ satisfies a second order differential equation of hypergeometric type and it can in general be expanded on the two solutions:
\begin{align}
G_{3}^{(1)}(x)&= x^{-\frac{4}{5}}(1-x)^{\frac{3}{5}}\; _{2} F_{1}\left(\frac{6}{5},-\frac{1}{5},-\frac{2}{5};x\right), \nonumber \\
G_{3}^{(2)}&= x^{\frac{3}{5}}(1-x)^{\frac{3}{5}}\; _{2} F_{1}\left(\frac{6}{5}, \frac{13}{5}, \frac{12}{5};x\right).
\end{align}
The small $x$ behavior $G_{3}^{(i)}(x)\sim x^{\Delta^{(i)}-\frac{4}{5}} (x<<1)$, associates the two solutions respectively to the identity channel $\Delta^{(1)}=0$ and to the $\Delta^{(2)}=\frac75$ channel.  In order to find how $Z_I$ and $Z_{II}$ are expressed in terms of the $G^{(i)}_3$, we study the behavior of $Z_{II}(x)$ in the limit for $x\to 0$. In this limit, the $Z_{II}$ is described by a $2$SLE$_{\kappa}$ process, with $\kappa=\frac{10}{3}$. This is a Bessel process  with fractal dimension \cite{Bauer2005}: 
\begin{equation}
\label{besselfra}
d_{f}=2+2(\Delta^{(i)}-2\Delta_{\varepsilon})-\frac{4}{\kappa}=2+ 2\Delta^{(i)}-\frac75.
\end{equation}
If we impose the non-recurrent condition for the Bessel process, we have
\begin{equation}
\label{nonrec} 
d_{f}>2,
\end{equation}
which implies:
\begin{equation}
Z_{II}(x)\propto G_{3}^{(2)}(x).
\end{equation}
One observes also that, as $Z_{II}(1-x)$ corresponds to $Z_{I}(x)$ with the boundary conditions $1$ and $23$ interchanged, 
one expects  $Z_{I}(x)=\beta Z_{II}(1-x)$, where $\beta$ is some constant. The fact that $\beta\neq 1$  originates from the fact 
that the condition $1$ and $23$ are not symmetric. The constant $\beta$ can be determined by imposing that the total partition 
function $Z(x)=Z_{I}(x)+Z_{II}(x)\sim x^{-\frac45} z(x)$, where $z(x)$ is polynomial in $x$. Using the monodromy properties of the 
hypergeometric function, one finds:
\begin{equation}
\label{beta}
\beta=-\frac{\Gamma(-\frac15)\Gamma(\frac65)}{\Gamma(-\frac75) \Gamma(\frac{12}{5})}\sim 1.61803
\end{equation} 
The crossing probability on the half plane are therefore:
\begin{align}
\label{Q=3CP}
{\cal P}_h(x)= & \frac{G^{(2)}_{3}(x)}{\beta\; G^{(2)}_3(1-x) +G^{(2)}_{3}(x)},\nonumber \\
 {\cal P}_v(x)= &\frac{\beta G^{(2)}_{3}(1-x)}{\beta\; G^{(2)}_3(1-x) +G^{(2)}_{3}(x)}.
\end{align}
We have verified that the above formulas coincide with the results in \cite{Flores_2017}.
Note that  $\mathcal{P}_{h}(\frac{1}{2})=\frac{1}{1+\beta}\neq \frac{1}{2}$, which makes the asymmetry between the boundary conditions manifest.

Let us consider now the $Z_4$ parafermionic spin model on the half plane, where no analytical prediction are known.
As in the previous case, we choose on the real line alternating boundary conditions that generate a single interface. 
For instance, in the following configuration:
\begin{center}
\begin{tikzpicture}
\draw[thick] (-3,0.5)--(13,0.5);
\draw[black,fill] (-2.5,0.5) circle [radius=0.1];
\draw[red,fill] (-1.5,0.5) circle [radius=0.1];
\draw[black,fill] (-0.5,0.5) circle [radius=0.1];
\draw[black,fill] (0.5,0.5) circle [radius=0.1];
\draw[green,fill] (-2.5,1.5) circle [radius=0.1];
\draw[red,fill] (-1.5,1.5) circle [radius=0.1];
\draw[black,fill] (-0.5,1.5) circle [radius=0.1];
\draw[blue,fill] (0.5,1.5) circle [radius=0.1];
\draw[blue,fill] (-2.5,2.5) circle [radius=0.1];
\draw[black,fill] (-1.5,2.5) circle [radius=0.1];
\draw[blue,fill] (-0.5,2.5) circle [radius=0.1];
\draw[blue,fill] (0.5,2.5) circle [radius=0.1];
\draw[red,fill] (-2.5,3.5) circle [radius=0.1];
\draw[black,fill] (-1.5,3.5) circle [radius=0.1];
\draw[green,fill] (-0.5,3.5) circle [radius=0.1];
\draw[green,fill] (0.5,3.5) circle [radius=0.1];
\draw (1,0.25)--(1,1)--(0,1)--(0,2)--(-1,2)--(-1,3.75);
\draw(1,0.25) node[below]{0};
\draw(-1,-0.25) node[below]{$S_i=1$ or $2$};
\draw[blue,fill] (1.5,0.5) circle [radius=0.1];
\draw[green,fill] (2.5,0.5) circle [radius=0.1];
\draw[green,fill] (3.5,0.5) circle [radius=0.1];
\draw[blue,fill] (4.5,0.5) circle [radius=0.1];
\draw[black,fill] (1.5,1.5) circle [radius=0.1];
\draw[green,fill] (2.5,1.5) circle [radius=0.1];
\draw[red,fill] (3.5,1.5) circle [radius=0.1];
\draw[black,fill] (4.5,1.5) circle [radius=0.1];
\draw[green,fill] (1.5,2.5) circle [radius=0.1];
\draw[blue,fill] (2.5,2.5) circle [radius=0.1];
\draw[red,fill] (3.5,2.5) circle [radius=0.1];
\draw[red,fill] (4.5,2.5) circle [radius=0.1];
\draw[black,fill] (1.5,3.5) circle [radius=0.1];
\draw[red,fill] (2.5,3.5) circle [radius=0.1];
\draw[red,fill] (3.5,3.5) circle [radius=0.1];
\draw[green,fill] (4.5,3.5) circle [radius=0.1];
\draw (5,0.25)--(5,1)--(3,1)--(3,3)--(1,3)--(1,3.75);
\draw(5,0.25) node[below]{x};
\draw(3,-0.25) node[below]{$S_i=3$ or $4$};
\draw[black,fill] (5.5,0.5) circle [radius=0.1];
\draw[red,fill] (6.5,0.5) circle [radius=0.1];
\draw[red,fill] (7.5,0.5) circle [radius=0.1];
\draw[black,fill] (8.5,0.5) circle [radius=0.1];
\draw[green,fill] (5.5,1.5) circle [radius=0.1];
\draw[red,fill] (6.5,1.5) circle [radius=0.1];
\draw[black,fill] (7.5,1.5) circle [radius=0.1];
\draw[black,fill] (8.5,1.5) circle [radius=0.1];
\draw[blue,fill] (5.5,2.5) circle [radius=0.1];
\draw[black,fill] (6.5,2.5) circle [radius=0.1];
\draw[blue,fill] (7.5,2.5) circle [radius=0.1];
\draw[blue,fill] (8.5,2.5) circle [radius=0.1];
\draw[red,fill] (5.5,3.5) circle [radius=0.1];
\draw[black,fill] (6.5,3.5) circle [radius=0.1];
\draw[green,fill] (7.5,3.5) circle [radius=0.1];
\draw[green,fill] (8.5,3.5) circle [radius=0.1];
\draw(7,-0.25) node[below]{$S_i=1$ or $2$};
\draw[green,fill] (9.5,0.5) circle [radius=0.1];
\draw[green,fill] (10.5,0.5) circle [radius=0.1];
\draw[green,fill] (11.5,0.5) circle [radius=0.1];
\draw[blue,fill] (12.5,0.5) circle [radius=0.1];
\draw[black,fill] (9.5,1.5) circle [radius=0.1];
\draw[red,fill] (10.5,1.5) circle [radius=0.1];
\draw[red,fill] (11.5,1.5) circle [radius=0.1];
\draw[black,fill] (12.5,1.5) circle [radius=0.1];
\draw[green,fill] (9.5,2.5) circle [radius=0.1];
\draw[blue,fill] (10.5,2.5) circle [radius=0.1];
\draw[red,fill] (11.5,2.5) circle [radius=0.1];
\draw[red,fill] (12.5,2.5) circle [radius=0.1];
\draw[black,fill] (9.5,3.5) circle [radius=0.1];
\draw[red,fill] (10.5,3.5) circle [radius=0.1];
\draw[red,fill] (11.5,3.5) circle [radius=0.1];
\draw[green,fill] (12.5,3.5) circle [radius=0.1];
\draw (9,0.25)--(9,1)--(10,1)--(10,2)--(7,2)--(7,3.75);
\draw(9,0.25) node[below]{1};
\draw(11,-0.25) node[below]{$S_i=3$ or $4$};

\end{tikzpicture}
\end{center} 
the boundary conditions on the real axis alternate between $S_i=1$ or $2$ and $S_i=3$ or $4$. 
We will assume that
\begin{itemize} 
\item the boundary changing condition fields are the parafermionic primaries $\varepsilon$, defined in Appendix \ref{ZN} 
with dimension $\Delta_{\varepsilon}$:
\begin{equation}
\label{bcc}
\Delta_{\varepsilon}=\frac{2}{N+2}
\end{equation} 
\item  the interfaces describe a multiple SLE process with \cite{Santachiara_2008}:
\begin{equation}
\kappa=4\;\frac{N+1}{N+2}, \quad N\geq 4
\end{equation} 
\end{itemize}
For $N=4$, $\Delta_{\varepsilon}=\frac{1}{3}$ and $\kappa=\frac{10}{3}$, and 
the fractal dimension of the $(12|34)$ interface is expected to be $1+\frac{\kappa}{8}=1+\frac{5}{12}$. As discussed before, 
this is consistent  with the numerical results, see Figure \ref{F2}. 
The $Z_{I}$ and $Z_{II}$ are related to the $Z_N$ parafermionic conformal blocks that contain  four $\varepsilon$ fields:
\begin{equation}
\label{CBpara}
G_N(x)=\left< \varepsilon(0) \varepsilon (x)\varepsilon(1)\varepsilon(\infty)\right>_{Z_N}.
\end{equation}
We show in Appendix \ref{computcb} that $G_N(x)$ satisfies, for $N\geq 4$, the third order linear differential 
equation (\ref{dif31})-(\ref{dif32}). The corresponding solutions, $G^{(1)}_N(x)$, $G^{(2)}_N(x)$ and $G^{(3)}_N(x)$ are 
associated to the three possible fields appearing in the fusion $\varepsilon\times \varepsilon$, see (\ref{idevare}) and (\ref{opephi}). 
The first terms in the small $x$ expansions of the $G^{(i)}_4(x)$ are given  in (\ref{solKZ}). The $Z_{I}$ and $Z_{II}$ can therefore be 
expanded on the $G^{(i)}_N$ basis, where: 
\begin{equation}
\label{asyG}
G_N^{(i)}(x)\sim \left< \varepsilon(0) \varepsilon (x) \phi^{(i)}(\infty)\right>\sim x^{-2\Delta_{\varepsilon}+\Delta^{(i)}},
\end{equation} where $\phi^{(i)}$, $i=1,2,3$, is one of the three fields in (\ref{opephi}) with  dimension $\Delta^{(1)}=0$, $\Delta^{(2)}=1+\frac{2}{N+2}$ and $\Delta^{(3)}=\frac{6}{N+2}$.
\noindent We can now use the same argument seen before, generalized to the non-Virasoro case \cite{fukusumi2017multiple}. 
In the limit $x\to 0$, $Z_{II}$ is described by a Bessel process  of fractal dimension (\ref{besselfra}). 
Comparing  (\ref{besselfra}) and (\ref{nonrec}) to (\ref{asyG}), the $G^{(i)}_N$ contributing to  $Z_{II}$  are 
the ones satisfying: \begin{equation}
\Delta^{(i)} >2\Delta_{\varepsilon}-\frac{2}{\kappa}.
\end{equation} 
For any $N\geq 4$, the above condition selects the fusion channels with $\Delta^{(2)}=1-\frac{2}{N+2}$ and $\Delta^{(3)}=\frac{2}{N+2}$. 
We conclude that $Z_{II}(x)$ is proportional to a linear combination of $G_N^{(2)}(x)$, and $G_N^{(3)}$ 
\begin{equation}
Z_{II}(x)\propto  G_N^{(2)}(x)+\alpha\; G_N^{(3)}(x).
\end{equation}
In the case of symmetric boundary conditions, such as the ones $(12|34|12|34)$ illustrated above, we can also set~:
\begin{equation}
Z_{I}(x)=Z_{II}(1-x).
\end{equation}
We obtain, for $N\geq 4$, the following expression:
\begin{align}
\label{z4CP}
{\cal P}_h(x)= & {G^{(2)}_N(x)+ \alpha\; G_N^{(3)}(x)\over G^{(2)}_N(x) + G^{(2)}_N (x)+\alpha\;\left(G_N^{(3)}(1-x)+G_N^{(3)}(1-x)\right)}\; \nonumber \\
 {\cal P}_v(x)= &{\cal P}_h(1-x)
\end{align}
We could not determine the constant $\alpha$. Analogously to the determination of $\beta$, see (\ref{beta}),  this requires to 
find the connection of a Fuchsian system. However, differently from the case $N=3$, for $N\geq 4$ this is a non-rigid rank $3$
Fuchsian system and finding the connection of this system is a very hard problem, see the discussion in \cite{bhs17}.  
We let  $\alpha$ undetermined and  we use it as a fit parameter  in the comparison with Monte-Carlo results.

\subsection{Measurements}
\label{measurements}

We present the results for crossing probabilities taken for a rectangle of ratio $r$ for the $Z_3$ and $Z_4$ spin models. 
By means of conformal map, see for instance \cite{Blanchard_2013}, one finds the relation between $r$ and $x$ appearing 
in the previous formula:
\begin{equation}
\label{rvsx}
r= \frac{K(x)}{K(1-x)}\; ,
\end{equation}
where  $K(x)$ is the complete elliptic integral of the first kind.

We give here some details on the measurements. For the $Z_4$ models, very 
efficient cluster algorithms are available \cite{WD}. We have just to keep the cluster changes compatible with the boundary 
conditions and this slows down the updates and limits the size we can simulate. For example, if during an update, a cluster is touching 
a boundary at the position $i$ with the condition $S_i =1$ corresponding to $\sigma_i = + $ and $\tau_i = +$, then this update can not 
be implemented. Indeed, any update\footnote{The update is done by considering separately an update of the $\sigma's$ or 
the $\tau's$, see \cite{WD} for details.} would change the sign of $\sigma_i$ or $\tau_i$ which would not preserve 
the boundary condition. We expect therefore non negligible finite size corrections to the measurements. 

For each value of $r$ and for each size $L_h$, we first made measurements of the autocorrelation time $\tau(L_h,r)$. 
Next we measure the probabilities $P_h$ and $P_v$ during a MC time $10^6 \tau(L_h,r)$ corresponding 
to  $10^6$ independent configurations. We performed measurements for $r=1, 1.2, 1.5 , 2.0$. For each value of $r$, 
we measure for $L_h = 10, 20, 40, 80, 160, 320$ and $640$. For $r=1.5$ we also simulated $L_h=1280$. 

Note that on the square lattice, for a given configuration, we can have the condition 
$P_h = P_v=0$. This corresponds to configuration with a corner. For example, 
the following configuration corresponding to $L_h = L_v =5$ with fixed boundary conditions,
$S_i=1$ on top and bottom (showed as filled black circles) and $S_i =2,3,4$ on left and right (showed as filled red, blue or green circles) :
\begin{center}
\begin{tikzpicture}
\draw[black] (0.5,0.5) circle [radius=0.1];
\draw[black,fill] (1.5,0.5) circle [radius=0.1];
\draw[black,fill] (2.5,0.5) circle [radius=0.1];
\draw[black,fill] (3.5,0.5) circle [radius=0.1];
\draw[black] (4.5,0.5) circle [radius=0.1];
\draw[red,fill] (0.5,1.5) circle [radius=0.1];
\draw[red,fill] (1.5,1.5) circle [radius=0.1];
\draw[blue,fill] (2.5,1.5) circle [radius=0.1];
\draw[black,fill] (3.5,1.5) circle [radius=0.1];
\draw[red,fill] (4.5,1.5) circle [radius=0.1];
\draw[blue,fill] (0.5,2.5) circle [radius=0.1];
\draw[black,fill] (1.5,2.5) circle [radius=0.1];
\draw[black,fill] (2.5,2.5) circle [radius=0.1];
\draw[blue,fill] (3.5,2.5) circle [radius=0.1];
\draw[green,fill] (4.5,2.5) circle [radius=0.1];
\draw[red,fill] (0.5,3.5) circle [radius=0.1];
\draw[black,fill] (1.5,3.5) circle [radius=0.1];
\draw[black,fill] (2.5,3.5) circle [radius=0.1];
\draw[red,fill] (3.5,3.5) circle [radius=0.1];
\draw[blue,fill] (4.5,3.5) circle [radius=0.1];
\draw[black] (0.5,4.5) circle [radius=0.1];
\draw[black,fill] (1.5,4.5) circle [radius=0.1];
\draw[black,fill] (2.5,4.5) circle [radius=0.1];
\draw[black,fill] (3.5,4.5) circle [radius=0.1];
\draw[black] (4.5,4.5) circle [radius=0.1];
\draw (0.3,1.2) -- (2.8,1.2);
\draw (2.8,1.2) -- ( 2.8, 1.8);
\draw ( 2.8, 1.8) -- (0.8, 1.8);
\draw (0.8, 1.8) -- (0.8, 3.8);
\draw (0.8, 3.8) -- (0.3, 3.8);
\draw (4.7, 3.8) -- (3.2,3.8);
\draw (3.2,3.8) -- ( 3.2, 2.2);
\draw ( 3.2, 2.2) -- (4.2, 2.2);
\draw (4.2, 2.2) -- (4.2, 1.2);
\draw (4.2, 1.2) -- (4.7,  1.2);
\draw (1.2, 0.3) -- (1.2,0.8);
\draw (1.2, 0.8) -- ( 3.2, 0.8);
\draw ( 3.2, 0.8) -- (3.2, 1.8);
\draw (3.2, 1.8) -- (3.8, 1.8);
\draw (3.8, 1.8) -- (3.8,  0.3);
\draw (1.2, 4.7) -- (1.2,2.2);
\draw (1.2, 2.2) -- ( 2.8, 2.2);
\draw ( 2.8, 2.2) -- (2.8, 4.2);
\draw (2.8, 4.2) -- (3.8, 4.2);
\draw (3.8, 4.2) -- (3.8,  4.7);
\end{tikzpicture}
\end{center}     
White circles on the upper/lower and left/right corners correspond to non connected spins.
For this particular configuration, we see that each border is connected to a cluster but no cluster is crossing the lattice.          
In practice, we observe that there is a very small number of such configurations 
and this number (or the proportion of this number) decrease with the linear size. To 
take in account these few configurations, we define 
the number of crossings as in (\ref{defP}).

\subsubsection{$(1|23)$ Potts spin interface}

We first present results for the crossing probabilities for the $Q=3$ Potts model with the conditions $(1|23)$. In Tab.~\ref{PyoPx}, we show the results 
for ${\cal P}_h(r)/{\cal P}_v(r)$  as a function of $r=L_h/L_v$. In this table, the first result is the prediction from (\ref{Q=3CP}) and the second result 
is the estimate for our numerical data. We performed simulations for $L_h \leq 640$ for $r \geq 1$ and for $L_v \leq 640$ for $r \leq 1$. 

\begin{figure}
\begin{center}
\epsfxsize=240pt\epsfysize=200pt{\epsffile{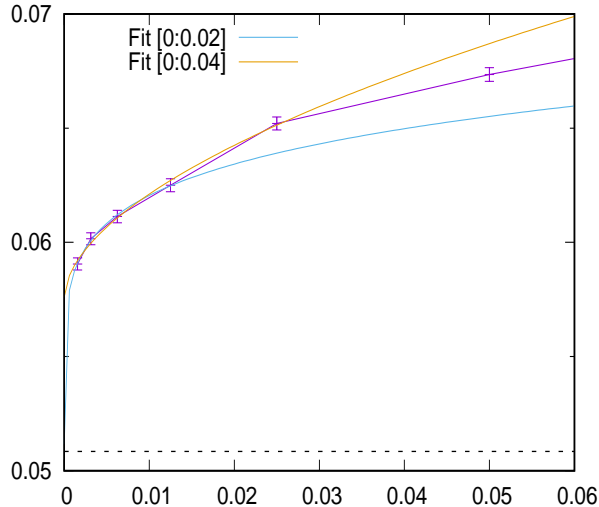}}
\end{center}
\caption{
${\cal P}_h(r)/{\cal P}_v(r)$ as a function of ${1/L_h}$ for the conditions $(1|23)$  for the $Q=3$ Potts model and for $r=1.5$.
\label{FP1}
}
\end{figure}

In practice, we observe that there are strong finite size corrections. We show an example for $r=1.5$ in Fig.~\ref{FP1} in which 
we plot ${\cal P}_h(r)/{\cal P}_v(r)$ vs. $1/L_h$. The value is expected to converge in the zero limit to the value $0.0508417$. 
Taking a fit of the form $a+b/L_h^c$, with data in the range $[0:0.02]$, we obtain $ a = 0.051 (9)$ which is compatible with the expected 
value. If we took a fit in the range $[0:0.04]$, then we obtain $a = 0.058 (1)$ which is close to the expected value but not any more compatible.
We also found that for all the values of $r$, the exponent $c$ is always very small, making a precise determination difficult. 

\begin{table}[h]
\centering
\begin{tabular}{ | c || c  |  c | c | c | c| c | c | c | c | c|} 
\hline
 $ r $                    &      From (\ref{Q=3CP})   & ${\cal P}_h(r)/{\cal P}_v(r)$ \\
\hline
  2.0                      &    0.00542833&  0.0059 (3) \\  %
  1.5                      &    0.0508417&  0.051  (9) \\ 
  1.1                      &    0.354786 &   0.373  (6) \\ 
  1.0                      &    0.618034 &   0.655  (6) \\ 
  1/1.1=0.909091  &    1.07661  &    1.19 (6)  \\ 
  1/1.5 = 0.666667&    7.51285  &     8.2 (1)   \\ 
  1/2 = 0.5             &   70.3653   &   75 (6)        \\ 
\hline
\end{tabular}
\caption{${\cal P}_h(r)/{\cal P}_v(r)$  in function of $r=L_v/L_h$. 
}
\label{PyoPx}
\end{table}

\subsubsection{$Z_4$ spin model}

\begin{figure}
\begin{center}
\epsfxsize=225pt\epsfysize=200pt{\epsffile{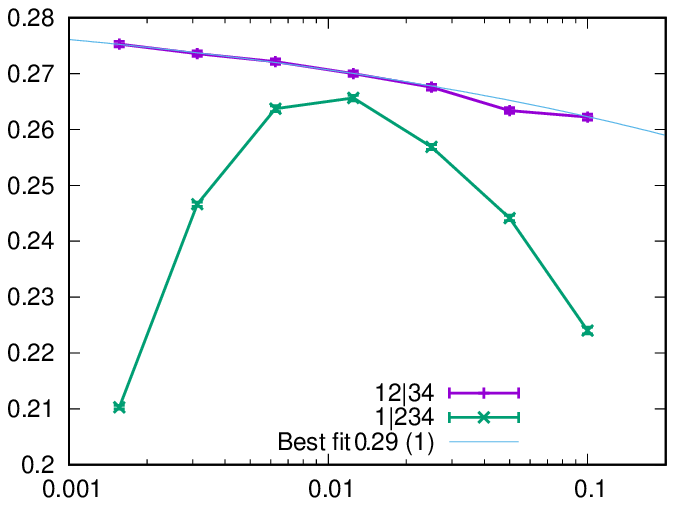}}
\epsfxsize=225pt\epsfysize=200pt{\epsffile{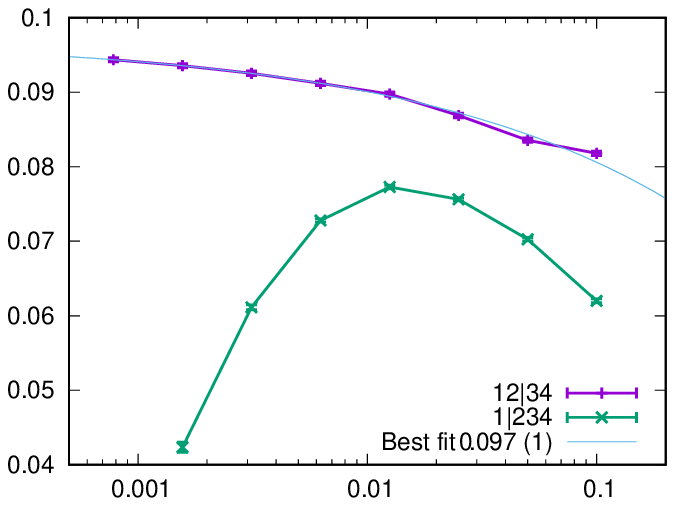}}
\epsfxsize=225pt\epsfysize=200pt{\epsffile{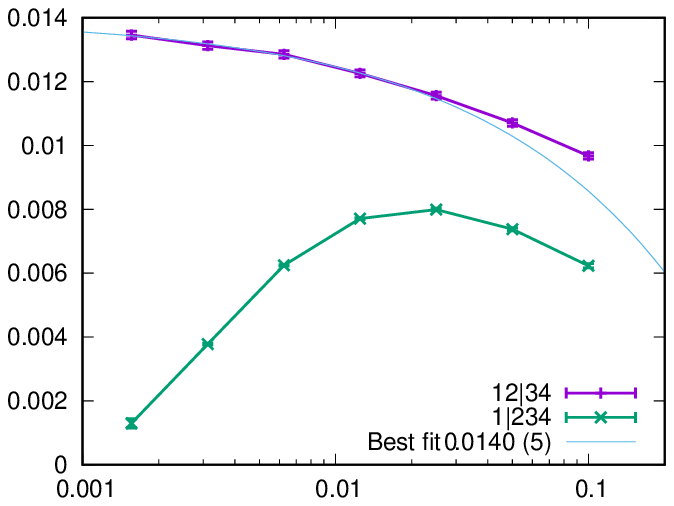}}
\end{center}
\caption{
${\cal P}_v(r)$ as a function of ${1/L_h}$ for the conditions $(1|234)$ and $(12|34)$. $r=1.2$ on the upper left, 
$r=1.5$ on the upper right and $r=2.0$ on the bottom.
\label{F1}
}
\end{figure}

In Fig.~\ref{F1}, we show our numerical results for ${\cal P}_v(r)$, the probability of having 
a cluster  crossing from the top to the bottom, as a function of $1/L_h$ for the condition $(1|234)$ 
and $(12|34)$. The figure contains data for $r=1,2$ in the upper left panel, for $r=1.5$ in the upper right 
panel and for $r=2.0$ in the lower panel.

We first consider the data for the boundary condition  $(1|234)$. In that case, for both aspect ratios, we observe 
that ${\cal P}_v(r)$ first increases as a function of $L_h$ and then falls down. 
The fact that it does not converge  is consistent with our previous observation that these boundary conditions are not the conformal 
boundary conditions, as discussed in Section (\ref{z4inter}), see Figure \ref{F2}.

Next, we consider the case with the boundary condition $(12|34)$. In order to take in account finite size effects,
we consider a fit to the form 
\beq
\label{fitP}
{\cal P}_h (L_h,r)= {\cal P}_h(\infty,r) +  \alpha(r) L_h ^{-\beta(r)} \; .
\eeq
In Fig.~\ref{F1}, we also show a fit of this for the boundary condition $(12|34)$. It converges nicely but the asymptotic limit 
still depends on the smallest value we keep for $L_h$. For instance for $r=1.5$, we obtained, for the following choices 
of smallest value $L_h$~:
\bea
L_h \geq 20  \rightarrow \beta(1.5) \simeq 0.46 (3) \; \; ; \; \; {\cal P}_h(\infty,1.5)  = 0.096 (1) \\
L_h \geq 40  \rightarrow \beta(1.5) \simeq 0.46 (7) \; \; ; \; \; {\cal P}_h(\infty,1.5)  = 0.096 (1) \\
L_h \geq 80  \rightarrow \beta(1.5) \simeq 0.29 (4) \; \; ; \; \; {\cal P}_h(\infty,1.5)  = 0.098 (1) \\
L_h \geq 160  \rightarrow \beta(1.5) \simeq 0.37 (1) \; \; ; \; \; {\cal P}_h(\infty,1.5)  = 0.097 (1) 
\eea
By extrapolation, we determine 
\beq
\label{z4data}
{\cal P}_h(\infty,1.2)  =  0.29 (1)  \; \; ; \; \; {\cal P}_h(\infty,1.5)  =  0.097 (2)  \; \; ; \; \;  {\cal P}_h(\infty,2.0)  =  0.0135 (10)  \; .
\eeq
We have also run measurements for $r=1$. For the condition $(12|34)$, we have a symmetry which ensure that ${\cal P}_h(\infty,1) = {\cal P}_v(\infty,1)$. 

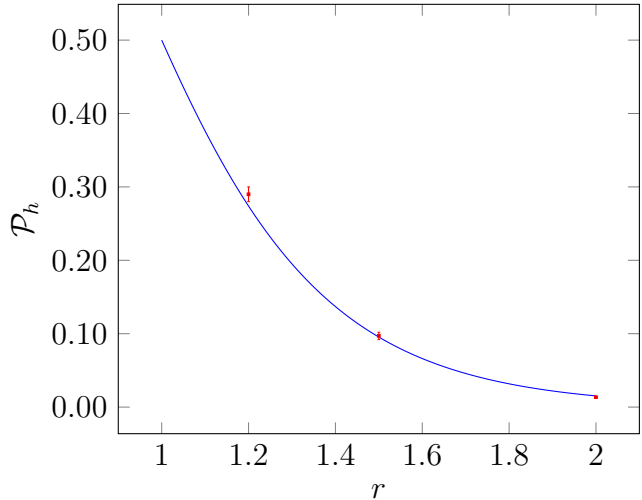
\begin{figure}
\centering
\begin{tikzpicture}
\begin{axis}[
	legend cell align=center,
	yticklabel style={/pgf/number format/.cd,fixed zerofill,precision=2},
	xlabel={$r$},
	ylabel={$\mathcal{P}_{h}$},
	legend pos=north west]
	]
\addplot+[smooth,no marks] 
     table[x=x,y=y]{
  x   y 	
  1 0.5
 1.01 0.486828 
 1.02 0.473801
 1.03 0.460932
 1.04 0.448232 
 1.05 0.435712 
 1.06 0.42338 
 1.07 0.411246 
 1.08 0.399317 
 1.09 0.3876 
 1.1 0.376102 
 1.11 0.364827 
1.12 0.353779 
1.13 0.342963 
1.14 0.332381 
1.15 0.322035 
1.16 0.311927 
1.17 0.302059 
1.18 0.29243 
1.19 0.28304 
1.2 0.273889 
1.21 0.264975 
1.22 0.256298 
1.23 0.247854 
1.24 0.239642 
1.25 0.23166 
1.26 0.223903 
1.27 0.21637 
1.28 0.209057 
1.29 0.20196 
1.3 0.195075 
1.31 0.188398 
1.32 0.181926 
1.33 0.175654 
1.34 0.169578 
1.35 0.163694 
1.36 0.157996
1.37 0.152481 
1.38 0.147145 
1.39 0.141982 
1.4 0.136989 
1.41 0.13216 
1.42 0.127492 
1.43 0.122979
1.44 0.118618
1.45 0.114405
1.46 0.110335
1.47 0.106403
1.48 0.102606
1.49 0.0989399
1.5 0.0954003
1.51 0.0919834
1.52 0.0886854
1.53 0.0855025
1.54 0.0824311
1.55 0.0794676
1.56 0.0766085
1.57 0.0738504
1.58 0.0711899
1.59 0.0686238
1.6 0.066149
1.61 0.0637623
1.62 0.0614609
1.63 0.0592418
1.64 0.0571022
1.65 0.0550394
1.66 0.0530507
1.67 0.0511336
1.68 0.0492855
1.69 0.0475041
1.7 0.0457871
1.71 0.0441321
1.72 0.0425369
1.73 0.0409996
1.74 0.0395179
1.75 0.03809
1.76 0.0367139
1.77 0.0353878
1.78 0.0341098
1.79 0.0328783
1.8 0.0316916
1.81 0.030548
1.82 0.029446
1.83 0.0283841
1.84 0.0273609
1.85 0.0263749
1.86 0.0254248
1.87 0.0245093
1.88 0.0236271
1.89 0.0227771
1.9 0.021958
1.91 0.0211687
1.92 0.0204081
1.93 0.0196753
1.94 0.0189691
1.95 0.0182886
1.96 0.0176328
1.97 0.017001
1.98 0.016392
1.99 0.0158053
2. 0.0152398
};
\addplot+[red,only marks, mark size = 0.5pt, error bars/.cd,y dir=both,y explicit] 
    table[x=x,y=y,y error = error] {
  x   y 	error
1.2	0.29	0.01
1.5	0.097	0.005
2	0.0135	0.001
};
\end{axis}
\end{tikzpicture}\caption{The (blue) line is the crossing probability (\ref{z4CP}) as a function of the rectangle ratio $r$, see (\ref{rvsx}). 
The value of the parameter $\alpha$ has been set by minimising the deviation from the numerical data from the prediction 
with $\alpha=0.325$. The (red) squares are the results for the asymptotic limit 
$\mathcal{P}_h(\infty,r)$ obtained from (\ref{fitP}).}\label{fig:compa}
\end{figure}

We can now compare the above results with the theoretical prediction (\ref{z4CP}) for $N=4$. This prediction is a function of a unknown constant $\alpha$.  
We thus perform a fit minimising, in function of $\alpha$, the relative difference between the measured data for $r=1.2, 1.5$ and $2.0$ and the prediction.
The minimal deviation is obtained for $\alpha= 0.325$ at which the (\ref{z4CP}) gives~:
\begin{align}
\label{z4theo}
&{\cal P}_h(1.2)  = 0.273889 \; \; ; \; \; {\cal P}_h(1.5)  = 0.0954003\quad {\cal P}_h(2)=0.0152398\; .
\end{align}
These values are in fact rather close to the measured values, see Fig.~\ref{fig:compa}. 
     
\section{Conclusion}

In this paper we have considered spin interfaces in the $Z_3$ ($Q=3$ state Potts) and in the  $Z_4$ parafermionic spin models. 
As shown in Figure \ref{F3} and in Figure \ref{F2}, the scaling behavior  of different spin interfaces is determined by 
two fractal dimensions, respectively $d_f=17/12$ and $d_f\sim 0.98$,  and $d_f=3/2$,  $d_f=17/12$. 
In particular the dimension $d_f=17/12$ was obtained for $Z_4$ spin models in previous works by using the SLE/ECFT correspondence, 
whose validity is thus supported by the numerical results presented here.  
We further investigated this correspondence by studying the crossing probabilities of spin interfaces. We compute the $Z_N$ crossing 
probability, given in (\ref{Q=3CP}) for $N=3$ and in (\ref{z4CP}) for $N\geq 4$, where the extended symmetry starts to play 
an important role. Note that the formula  (\ref{z4CP}) is given in terms of a parameter $\alpha$ that we could not fix.  
The parafermionic conformal blocks entering the (\ref{z4CP}) have been computed by using the $SU(2)_N$ coset description 
of the parafermionic CFT, and in particular by solving the Kniznhik-Zamolodchikov system of rank 3. Full details are provided 
in the Appendices.  For the $Q=3$ spin Potts model, we have verified that the parafermionic conformal blocks coincides with 
the one of the Virasoro algebra and that the formula (\ref{Q=3CP}) is a  special case of the crossing formulas given in \cite{Flores_2017}. 
Our numerical data, shown in Table \ref{PyoPx} are in good agreement with the prediction (\ref{Q=3CP}). The comparison between 
the numerical results for $N=4$, and the prediction (\ref{z4CP}) is shown in (\ref{z4data})-(\ref{z4theo}) and in Figure \ref{fig:compa}. 
Taking into account the presence of finite size effects, the agreement between the numerical and theoretical prediction is good. 
 
\appendix

\section{$Z_N$ parafermionic theory} 
\label{ZN}

The $Z_N$ theories have central charge:
\begin{equation}
c_N= \frac{2(N-1)}{N+2}
\end{equation}
and describe critical theories with dihedral group symmetry $D_{N}=Z_N\otimes (Z_2)^{N}$ \cite{Zampara}. 
The Hilbert space of the  $Z_N$ theories is constructed from the representations of the parafermionic algebra. 
The representation modules can be split into two sectors  \cite{Dotsenko_2003}, the disorder and the charge sector, 
associated respectively to the $(Z_2)^N$ and to the $Z_N$ element of the group $D_{N}$.  Here we are interested in 
the charge sector, that contains   $\lfloor \frac{N}{2}\rfloor +1$ representation modules $\mathcal{V}_j$, where $\lfloor x\rfloor$ is 
the integer part of $x$:
\begin{align}
\text{Parafermionic primaries:}&\quad  \Phi^{\pm j} \left\{
\begin{array}{c l}	
     j=0,1,\cdots, \frac{N-1}{2}& (N\;\text{odd})\\
     j=0,\frac12,\cdots, \frac{N}{4}& (N\;\text{even})
\end{array}\right.
\end{align}
Each  module $\mathcal{V}_j$ is formed by the descendants of:
\begin{equation}
\Phi^{\pm j}, \quad j=0,1,\cdots, \frac{N-1}{2}\quad N\;\text{odd},\quad   j=0,\frac12,\cdots, \frac{N}{4}\quad N\;\text{even}
\end{equation}
that have  $Z_N$ charge $j$ (defined mod $N$) and conformal dimension:
\begin{equation}
\Delta_{j}=\frac{j(N-2 j)}{N(N+2)}.
\end{equation}
When $j=-j \; \mod{N}$, the two fields $ \Phi^{\pm j}$ coincide. 

\noindent Besides the Identity, $\text{Id}=\Phi^{0}$, there are  $\lfloor \frac{N}{2}\rfloor$ Virasoro primaries $\varepsilon^{(j)}$ 
with zero $Z_N$ charge. They appear in $\mathcal{V}_j$ as the lowest dimension descendant. Among these neutral fields, 
the field $\varepsilon= \varepsilon^{(1)}$  has the lowest dimension given in (\ref{bcc}). 

The above field is the one that was considered in \cite{Santachiara_2008,pisa07} as the boundary changing condition 
field generating the SLE interface. Accordingly we expect the crossing probability associated to these SLE interfaces 
to be given by the conformal block (\ref{CBpara}).

\noindent For $N=2$ (Ising) and $N=3$ (3-state Potts), one can show that $G_{N(=2,3)}(x)$ satisfies the same correlations 
functions as the one of the corresponding  Virasoro minimal model. In \cite{Santachiara_2008} this was explicitly 
shown for $N=3$ and  we provide below also the demonstration for $N=2$, missing in \cite{Santachiara_2008}.

\subsection{Ising case, $N=2$}
\label{Ising}

 In this case one has the $\mathcal{V}_0$ (Neveu-Schwartz sector) and $\mathcal{V}_{\frac{1}{2}}$ (Ramond sector) 
 representation module. We denote as $\left[\Phi^{0}\right]$ a general  (parafermionic) descendants field in identity 
 field module $\mathcal{V}_0$. The (para)fermionic current $\psi(z)$ satisfies the OPE 
\begin{equation}
\label{opez2}
\psi(z)\psi(0)=\frac{1}{z}+ 2\;z^2 \; T^{Z_2}(0),
\end{equation} 
where $ T^{Z_2}$ is the stress-energy tensor
\begin{equation}
 T^{Z_2}(z)\left[\Phi^{0}\right](0)=\sum_{n=-\infty}^{\infty}\; \frac{1}{z^{n+2}}\;L^{Z_2}_{n}\;\left[\Phi^{0}\right](0)
\end{equation} whose modes $L^{Z_2}_{n}$ generate a $c=\frac12$ Virasoro algebra. The $\psi$ modes 
acting in $\mathcal{V}_0$ can be expressed as:
\begin{equation}
\label{modesz2}
A_{-\frac{1}{2}+n} \left[\Phi^{0}\right](0)=\frac{1}{2\pi i}\oint_{\mathcal{C}_0} d \;z \;z^{n-1}\; \psi(z) \left[\Phi^{0}\right](0),
\end{equation}  
where $\mathcal{C}_z$ is a contour encircling $z$. The current modes satisfy the (anti-)commutation relation:
\begin{equation}
\label{freefer}
\{A_{-\frac{1}{2}+n}, A_{-\frac{1}{2}+m}\}=\delta_{n+m,1}, 
\end{equation} 
that is a direct consequence of the fact that $\psi(z)$ is a free fermion. 
As previously said, $\varepsilon$ appears as a parafermionic descendant. In the case $N=2$ it appears in the identity module:
\begin{equation}
\varepsilon = A_{-\frac12}\; \Phi^{0}.
\end{equation}
Note that, setting $n=0,m=0$ in (\ref{freefer}), one gets:
\begin{equation}
A_{-\frac{1}{2}}A_{-\frac{1}{2}}\Phi^{0}=A_{-\frac{1}{2}}\varepsilon=0.
\end{equation}
From (\ref{modesz2}) and using standard contour integral manipulations, one finds that:
\begin{align}
&\frac{1}{(2\pi i)^2}\;\oint_{\mathcal{C}_0} d\; z \;\oint_{\mathcal{C}_z}\; (z)^{n-1}\;(z')^{m-1} (z-z')^{-2} \;\psi(z)\;\psi(z') \left[\Phi^{0}\right](0)=\nonumber \\
&=\sum_{k=0}^{\infty}\; (-1)^k \binom{-2}{k}\; \left( A_{-\frac12+n-2-k}A_{-\frac12+m+k}+ A_{-\frac12+m-2-k}A_{-\frac12+n+k}\right)
\end{align}
Using (\ref{opez2}) and the Cauchy theorem, one obtains the following relations between the $\psi$ and $T^{Z_2}$ modes:
 \begin{align}
&\sum_{k=0}^{\infty}\; (-1)^k \binom{-2}{k}\; \left( A_{-\frac12+n-2-k}A_{-\frac12+m+k}+ A_{-\frac12+m-2-k}A_{-\frac12+n+k}\right)=\nonumber \\
&\frac{(m-1)(m-2)}{2}\delta_{n+m,3}+2\; L^{Z_2}_{n+m-3}.
\end{align}
By setting $n=1,m=1$ and $n=1,m=0$ in the above formula and using the commutations (\ref{freefer}), one obtains:
\begin{align}
&(L_{-1}^{Z_2})^2\varepsilon = \left(A_{-\frac32}A_{\frac12}+2 A_{-\frac52}A_{+\frac32}\right)A_{-\frac32}A_{\frac12}\varepsilon =2A_{-\frac52}A_{\frac12}\varepsilon\nonumber \\
& L_{-2}^{Z_2}\;\varepsilon=\left(\frac12 A_{-\frac32}A_{-\frac12}+\frac32 A_{-\frac52}A_{\frac12}\right)\varepsilon=\frac32\;A_{-\frac52}A_{\frac12}\varepsilon.
\end{align} 
This implies:
\begin{equation}
\left((L_{-1}^{Z_2})^2-\frac{4}{3}L^{Z_2}_{-2}\right)\varepsilon=0,
\end{equation}
that coincides with the second-order null vector condition of the minimal model at $c=\frac12$.
\section{Computation of $G_{N}(x)$ for $N\geq 4$} 
\noindent We show now how to compute $G_N(z)$ for any $N$ by using the coset $\frac{SU(2)_{N}}{U(1)_N}$ 
description of the charge sector. In order to fix conventions and notations, let us first briefly review the 
$U(1)_N$ and $SU(2)_N$ CFTs. 

The $U(1)_N$ theory is a CFT with a current  $J= i\partial \phi$ of conformal dimension one, 
where $\phi$ is a free Gaussian field. The current modes $J_n$ form the Heisenberg algebra:
\begin{equation}
[J_{n},J_{m}]= n\delta_{n+m,0}.
\end{equation}
The stress energy-tensor:
\begin{equation}
T^{U(1)}=\frac{1}{2}:J\;J:,
\end{equation}
forms a Virasoro algebra with central charge:
\begin{equation}
c^{U(1)}=1.
\end{equation}
The $U(1)_N$ primary fields are the vertex fields $e^{i \frac{m}{\sqrt{N}}\phi}$ labeled by $m\in \mathbb{Z}/2$, 
defined modulo $N$, that have conformal dimension $\Delta^{U(1)}_{m}$:
\begin{equation}
\Delta^{U(1)}_{m}=\frac{m^2}{N}.
\end{equation}

\noindent The $SU(2)_N$ theory is based on the algebra formed by three conserved current $J^{a}$, $a=0,+,-$ 
of conformal dimension one:
\begin{equation}
[J^{a}_{n},J^{b}_{m}]=f^{a b}_{c} J^{c}_{n+m}+\frac{N n}{2} q^{a b}\delta_{n+m} 
\end{equation}
with the structure constants $f^{\pm, \mp}_{0}=\pm 2$, $f^{0, \pm}_{0}=\pm 2$ and
the Killing form $q^{a,b}$ is  $q^{0,0}=1$, $q^{+, -}=q^{-,+}=2$.  The stress-energy tensor is given by:
\begin{equation}
\label{virsu2}
T^{SU(2)}=\frac{1}{N+2}q_{a,b} :J^{a}J^{b}:,
\end{equation}
where $::$ denotes the regular part and it generates a Virasoro algebra with central charge:
\begin{equation}
c^{SU(2)_N}=\frac{3 N}{N+2} \; .
\end{equation}
The  representation module $\mathcal{V}^{SU(2)}_{l}$, $l\in \frac{\mathbb{N}}{2}$, contains the 
Virasoro primary fields $\phi^{j}_{m}$  that have conformal dimension:
\begin{equation}
\Delta^{SU(2)}_{j}=\frac{j(j+1)}{N+2}
\end{equation}
and transform under the action of $J^{a}_{0}$ as the $\it{su}(2)$ Lie algebra weights $|l,m\rangle$, $m=-l,-l+1,\cdots,l$. 
In particular, the $SU(2)$ Ward identities  are encoded in the following OPE:
\begin{equation}
\label{Ward}
J^{a}(x) \phi^{j}_{m}(0)= \frac{1}{x}\;\sum_{m'=-j}^{j}(t^{a})^{(j)}_{m,m'}\;\phi^{j}_{m'}(0)+\text{Regular terms},
\end{equation}
where $(t^{a})^{(j)}_{m,m'}$ is the $(2j+1)\times (2j+1)$ matrix in the representation $j$. 
As $\left< J(x) \prod_{i}\phi^{j_i}_{m_i}\right>\sim x^{-2}$ for $x\to \infty$, the above OPE implies:
\begin{equation}
\label{Ward2}
\sum_{i} (t^{a})^{(j_i)}_{m_i,m'_i}\left< \prod_{i}\phi^{j_i}_{m'_i}(x_i)\right>=0
\end{equation} 
From (\ref{virsu2}) one can show also  that the $SU(2)$ conformal blocks satisfy the Kniznhik-Zamolodchikov equations \cite{KNIZHNIK198483}:
\begin{equation}
\label{KZ}
\left[\partial_{x_i}+\sum_{j\neq i}\frac{q_{a,b}}{z-z_i} (t^{a})^{(j)}(t^{b})^{(i)}\right]\left<\prod_{l}\phi^{l_i}_{m_i} \right>=0
\end{equation}

\noindent In the coset  construction $\frac{SU(2)_{N}}{U(1)_N}$, the stress-energy 
tensor $T^{Z_N}$ of $Z_N$ satisfies the relation 
\begin{equation}
T^{SU(2)}= T^{U(1)}+T^{Z_N},
\end{equation} 
and the primaries fields of the three CFTs are related by:
\begin{equation}
\label{cosetid}
\phi^{j}_{\pm j} = \Phi^{j}\; e^{i \frac{\pm j}{\sqrt{N}} \phi},\quad \phi^{j}_{0}=\varepsilon^{(j)}
\end{equation}
In particular we have for the $\varepsilon =\varepsilon^{(1)}$ fields in (\ref{CBpara}):
\begin{equation}
\label{idevare}
\varepsilon = \phi^{1}_0
\end{equation}
\subsection{$SU(2)_N$ conformal blocks and Khniznik-Zamolodchikov equation}
\label{computcb}
By using (\ref{idevare}), we will write the function $G_N(z)$ in (\ref{CBpara}) as: 
\begin{equation}
\label{Gsu2}
G_N(x) =\left< \phi^{1}_{0}(0)\phi^{1}_{0}(x)\phi^{1}_{0}(1)\phi^{1}_{0}(\infty)\right>,
\end{equation}
and we will solve the corresponding equation (\ref{KZ}).

 \subsubsection{Coulomb gas representation of the rank 3 KZ system}
The solution of the KZ system can be given in terms of Coulomb gas integrals \cite{DOTSENKO1991547,CF_WZW}. 
\noindent Take the  more general function 
\begin{equation}
G^{\{m_i\}}(x)=\left<\phi^{1}_{m_1}(0)\phi^{1}_{m_2}(x)\phi^{1}_{m_3}(1)\phi^{1}_{m_4}(\infty) \right>,\quad \sum_i m_i=0.
\end{equation} This function lives in the $SU(2)$ singlet of the fusion $1\otimes 1\otimes 1\otimes 1$  of 
four spin-1 representations. In this case the invariant space has dimension $3$, i.e. it is spanned by 
three independent tensors $T^{(l)}_{\{m_i\}}$ with $l=0,1,2$:
\begin{equation}
T^{(l)}_{\{m_i\}}=\frac{1}{2l +1}\sum_{m=-l}^{l} \left<l,m|1,m_1; 1, m_2\right>\left<l,-m|1, m_3; 1, m_4\right>.
\end{equation} 
where $\left<j_3,m_3|j_1, m_1; j_2, m_2\right>$ are the $SU(2)$ Clebsh-Gordan coefficients. As usual for 
conformal blocks, specifying the primaries $\{m_i\}$ does not completely determine the function: we need to 
specify the quantum numbers $j$ and $m$ of the $\Phi^{j}_{m}$ in the internal channel. 
In the following we will use the basis of \cite{CF_WZW}. Note that we could equivalently choose the one in \cite{DOTSENKO1991547}.
In the tensor basis $T^{(l)}_{\{m_i\}}$,  we have:
\begin{equation}
\label{expsing}
G^{\{m_i\}}(z)=\left[\sum_{l=0}^{2} T^{(l)}_{\{m_i\}} g^{(l)}(z)\right] \; ,
\end{equation}
where the functions $f^{(l)}(z)$ are expressed via Coulomb gas integrals, see below. Using in (\ref{expsing}) 
the Clebsh-Gordan coefficients
\begin{eqnarray}
\left< l=0 ,m=0| 1,m_1=0;1,m_2=0\right> &=&-\sqrt{\frac{1}{3}},\nonumber \\ 
\left< l=2 ,m=0| 1,m_1=0;1,m_2=0\right> &=&\sqrt{\frac{2}{3}}\nonumber \\
\left< l=1 ,m=0| 1,m_1=0;1,m_2=0\right> &=& 0,
\label{CGcoeff}
\end{eqnarray}
one can see that $G_N(x)$ gets contribution from $g^{(0)}(x)$ and $g^{(2)}(x)$, 
that are given in \cite{CF_WZW}\footnote{We use different notation from \cite{CF_WZW}. For instance, 
the functions $g^{(k)}(x)$ here corresponds to the $f^{(k+1)}(\eta)$ given in Eq. (3.6) of \cite{CF_WZW}}:
\begin{eqnarray}
\label{definf}
g^{(i)}(x) &=& z^{\frac{2}{N+2}} (1-x)^{\frac{2}{N+2}}\int_{\mathcal{C}} d u_{1} d u_2 \;f^{(i)} (u_1,u_2,x),\quad i=0,2,
\end{eqnarray}
where \begin{align}
f^{(0)} (u_1,u_2,x)=& u_1^{-\frac{2}{N+2}-1} u_2^{-\frac{2}{N+2}-1} (u_1-1)^{-\frac{2}{N+2}}(u_2-1)^{-\frac{2}{N+2}}\;\nonumber \\& (u_1-x)^{-\frac{2}{N+2}-1}(u_2-x)^{-\frac{2}{N+2}-1}(u_1-u_2)^{\frac{2}{N+2}},\nonumber \\
f^{(2)} (u_1,u_2,z)=& u_1^{-\frac{2}{N+2}} u_2^{-\frac{2}{N+2}} (u_1-1)^{-\frac{2}{N+2}-1}(u_2-1)^{-\frac{2}{N+2}-1}\;\nonumber \\& (u_1-x)^{-\frac{2}{N+2}}(u_2-x)^{-\frac{2}{N+2}}(u_1-u_2)^{\frac{2}{N+2}}.
\end{align} 

\noindent In (\ref{definf}), $\mathcal{C}$ is a closed contour in the Riemann surface associated to the multi-valued functions $f^{(i)}(u_1,u_2,x)$. 
\noindent The integrals:
\begin{align}
I^{(i)}_{\mathcal{C}_{1}}(x) &=  x^{\frac{2}{N+2}+i} (1-z)^{\frac{2}{N+2}} \int_{0}^{x}\; d\; u_1\;\int_{0}^{u_1}\; d\; u_2 \;f^{(i)} (u_1,u_2,x) \nonumber \\
I^{(i)}_{\mathcal{C}_{2}}(x) &= x^{\frac{2}{N+2}+i} (1-z)^{\frac{2}{N+2}}\int_{0}^{x}\; d\; u_1\;\int_{1}^{\infty}\; d\; u_2 \;f^{(i)} (u_1,u_2,x)\nonumber \\
I^{(i)}_{\mathcal{C}_{3}}(x) &= x^{\frac{2}{N+2}+i} (1-z)^{\frac{2}{N+2}}\int_{1}^{\infty}\; d\; u_1\;\int_{1}^{u_1}\; d\; u_2\;f^{(i)} (u_1,u_2,x),\end{align} 
correspond to the conformal blocks that transform diagonally under the monodromy around the $x=0$ 
singularity and form a basis, the $s-$channel basis, on which the functions $g^{(i)}(x)$ can be expanded:
\begin{equation}
g^{(i)}(x)\to \{I^{(i)}_{\mathcal{C}_{1}}(x),I^{(i)}_{\mathcal{C}_{2}}(x),I^{(i)}_{\mathcal{C}_{3}}(x)\},
\end{equation}


\noindent  One can read the small $x$ asymptotic behavior of $I^{(i)}_{\mathcal{C}_{j}}(x)$:
\begin{align}
&I^{(0)}_{\mathcal{C}_{3}}(x) \sim x^{\frac{2}{N+2}+2}\left(1+\cdots\right), \quad I^{(2)}_{\mathcal{C}_{3}}(x)\sim x^{\frac{2}{N+2}}\left(1+\cdots\right)\nonumber \\
&I^{(0)}_{\mathcal{C}_{2}}(x) \sim x^{1-\frac{2}{N+2}}\left(1+\cdots\right), \quad I^{(2)}_{\mathcal{C}_{2}}(x)\sim x^{1-\frac{2}{N+2}}\left(1+\cdots\right)\nonumber \\
&I^{(0)}_{\mathcal{C}_{1}}(x) \sim x^{-\frac{4}{N+2}}\left(1+\cdots\right), \quad I^{(2)}_{\mathcal{C}_{1}}(x)\sim x^{-\frac{4}{N+2}+2}\left(1+\cdots\right)
\end{align}
The exponents
\begin{equation}
\label{expKZ}
\alpha_1=-\frac{4}{N+2}, \quad \alpha_2= 1-\frac{2}{N+2}, \quad \alpha_3=\frac{N}{N+2}
\end{equation}
correspond respectively to the fusion channels:
\begin{equation}
\label{opephi}
\phi^{1}_{0} \times \phi^{1}_{0}=\phi^{0}_{0}+J^{\pm }_{-1}\phi^{1}_{\mp 1}+\phi^{2}_{0},
\end{equation}
Notice that the fusion coefficient $ \phi^{1}_{0} \times \phi^{1}_{0}\to \phi^{1}_{0}$ vanishes with the 
corresponding Clebsh-Gordon coefficient, see (\ref{CGcoeff}).

\subsubsection{Solution of the rank 3 KZ by the Frobenious method }
We have seen that the solutions of the rank 3 system are expressed in terms of bidimensional integrals 
whose evaluation requires to compute double infinite sums of $_{3} F_{2}$ hypergeometric functions. 
We verified that the convergence of these sums is quite slow. 
A more efficient method is to use the Frobenious method to find a series expansion of the solution. 
We notice that this method for a Fuchsian system of rank 3 has been also applied in \cite{Gori_2018}. 

\noindent In our conventions the matrix appearing in (\ref{KZ}) have the form:
\begin{equation}
(t^{+})^{(1)} = \left[
\begin{array}{rrr}
0 & -i & 0 \\
0 & 0 & -2i \\
0 & 0 & 0
\end{array}
\right],
(t^{-})^{(1)} = \left[
\begin{array}{rrr}
0 & 0 & 0 \\
2i & 0 & 0 \\
0 & i & 0
\end{array}
\right]
(t^{0})^{(1)} = \left[
\begin{array}{rrr}
1 & 0 & 0 \\
0 & 0 & 0 \\
0 & 0 & -1
\end{array}
\right] \; .
\end{equation}
Using the above matrices, one can verify that the (\ref{KZ}) is a system of first-order differential equation 
that couples the conformal block (\ref{Gsu2}) with other conformal blocks containing four spin-$1$ representation. 
Using the Ward identities (\ref{Ward2}), one can show that the space of functions appearing in the KZ system is spanned
 by three independent functions. We take as a basis the following ones:
%
\begin{align}
a(x)&=\left< \phi^{1}_{-1}(0)\phi^{1}_{1}(x)\phi^{1}_{-1}(1)\phi^{1}_{1}(\infty)\right> \\
b(x)&=\left< \phi^{1}_{0}(0)\phi^{1}_{0}(x)\phi^{1}_{-1}(1)\phi^{1}_{1}(\infty)\right>\\
G_N(x)&=\left< \phi^{1}_{0}(0)\phi^{1}_{0}(x)\phi^{1}_{0}(1)\phi^{1}_{0}(\infty)\right>
\end{align}

%


\noindent In the basis of functions $(a(x),b(x),G_N(x))$, the KZ equation (\ref{KZ}) takes the form:
\begin{equation}
\label{fuchs3}
\frac{N+2}{2}\partial_{x}F=\frac{1}{x}A_0 F+\frac{1}{x-1}A_1 F,
\end{equation}
where $F$ is defined by $(a(x),b(x),G_N(x))^{t}$ and 

\begin{equation}
A_0 = \frac{2}{N+2}\left[
\begin{array}{rrr}
-1 & 2 & 0 \\
0 & -1 & 2 \\
0 & 1 & 0
\end{array}
\right],\quad 
A_1 = \frac{2}{N+2}\left[
\begin{array}{rrr}
1 & 0 & 0 \\
-\frac12 & -1 & 0 \\
\frac12 & 0 & -2
\end{array}
\right] \; .
\end{equation}
The Fuschian system (\ref{fuchs3}) implies that the $G_N(x)$ satisfies a 3rd order differential equation of Fuchsian type:
\beq
\label{dif31}
\left(\partial^3_{x} +\frac{h_2(N,x)}{x(x-1)}\partial^2_{x}+\frac{h_1(N,x)}{x^2(x-1)^2}\partial_{x}+ \frac{h_0(N,x)}{x^2(x-1)^3} \right)G_{N}(x)=0,
\eeq 
with:
\begin{align}
\label{dif32}
h_{2}(N,x)&=\frac{2(-7-N+ (11+2 N) x )}{(N+2)},\nonumber \\
h_{1}(N,x)&=\frac{2(16+2 N-(72+17 N+N^2) x +(56+16N+N^2) x^2)}{(N+2)^2}\nonumber \\
h_{0}(N,x)&=\frac{4(24+3 N-(48+5 N) x +(24+4 N) x^2)}{(N+2)^3} \; .
\end{align}
The corresponding vector of solutions $G_N^{(i)}(z)$, $i=1,2,3$ can be given as an expansion around one of 
the branch singularities at $x=0,1$ and $x=\infty$. Expanding around the $x=0$ singularities, the solutions take the form
\begin{equation}
\label{serexp}
G_N^{(i)}(x)=x^{\alpha_i}\sum_{n=0}^{\infty}\;p^{(i)}_{n}(N)\;x^{n}.
\end{equation}
In the above equation $\alpha_i$ takes one of the values of the $A_{0}$ eigenvalues
\beq
A_{0}\sim \text{diag}[\alpha_{1},\alpha_2,\alpha_3],
\eeq 
that are given in (\ref{expKZ}). For each $\alpha_i$, the coefficients $p^{(i)}_n(N)$ in (\ref{serexp}) 
can be obtained very efficiently  by using the Frobenious method for Fuchsian equations.
For $N=4$, that is the value at which  we compare the analytical formula (\ref{z4CP}) to numerical simulations, we find:
\begin{align}
G_4^{(1)}(x)&=x^{-\frac{2}{3}}\left(1+\frac{2}{9}x^{2}+\frac{2}{9}x^{3}
+\frac{103}{486}x^{4}+\frac{49}{253}x^{5}+\cdots\right)\nonumber \\
G_4^{(2)}(x)&=x^{\frac{2}{3}} \left(1+\frac{2}{3}x+\frac{5}{9}x^{2}+\frac{40}{81}x^{3}+\frac{110}{243}x^{4}+\frac{308}{729}x^{5}+...\right),\nonumber \\
G_4^{(3)}(x)&=x^{\frac{1}{3}}\left(1+\frac{1}{2}x+\frac{11}{27}x^{2}+\frac{13}{36}x^{3}+\frac{161}{486}x^{4}+\frac{301}{971}x^{5}+...\right).
\label{solKZ}
\end{align}

\acknowledgments{{We thank Flavia Albarracin for contributing at an early stage of this project and Nina Javerzat and 
Jacopo Viti for useful discussions. YF acknowledges the support by the QuantiXLie Center of Excellence, 
a project co-financed by the Croatian Government and European Union 
through the European Regional Development Fund - the Competitiveness 
and Cohesion Operational Programme (Grant KK.01.1.1.01.0004). R.S. acknowledges the hospitality of the 
International Institute of Physics (Natal) where part of this work has been done. 
}}
\bibliography{bibparafermion_YF_v2}

\providecommand{\href}[2]{#2}\begingroup\raggedright\begin{thebibliography}{10}

\bibitem{PSS}
M.~{Picco}, R.~{Santachiara} and A.~{Sicilia}, \emph{{Geometrical properties of
  parafermionic spin models}},
  \href{http://dx.doi.org/10.1088/1742-5468/2009/04/P04013}{\emph{Journal of
  Statistical Mechanics: Theory and Experiment} (2009) P04013},
  [\href{http://arxiv.org/abs/0812.3526}{{\tt 0812.3526}}].

\bibitem{PS2}
M.~{Picco} and R.~{Santachiara}, \emph{{Critical interfaces of the
  Ashkin-Teller model at the parafermionic point}},
  \href{http://dx.doi.org/10.1088/1742-5468/2010/07/P07027}{\emph{Journal of
  Statistical Mechanics: Theory and Experiment} (2010) P07027},
  [\href{http://arxiv.org/abs/1005.0493}{{\tt 1005.0493}}].

\bibitem{Flores_2017}
S.~M. Flores, J.~J.~H. Simmons, P.~Kleban and R.~M. Ziff, \emph{A formula for
  crossing probabilities of critical systems inside polygons},
  \href{http://dx.doi.org/10.1088/1751-8121/50/6/064005}{\emph{Journal of
  Physics A: Mathematical and Theoretical} {\bf 50} (Jan, 2017) 064005}.

\bibitem{prs19}
M.~Picco, S.~Ribault and R.~Santachiara, \emph{{On four-point connectivities in
  the critical 2d Potts model}},
  \href{http://dx.doi.org/10.21468/SciPostPhys.7.4.044}{\emph{SciPost Phys.}
  {\bf 7} (2019) 44}.

\bibitem{saleur2020}
Y.~He, J.~L. Jacobsen and H.~Saleur, \emph{{Geometrical four-point functions in
  the two-dimensional critical $Q$-state Potts model: The interchiral conformal
  bootstrap}},  \href{http://arxiv.org/abs/2005.07258}{{\tt 2005.07258}}.

\bibitem{jps19two}
N.~Javerzat, M.~Picco and R.~Santachiara, \emph{{Two-point connectivity of
  two-dimensional critical Q-Potts random clusters on the torus}},
  \href{http://dx.doi.org/10.1088/1742-5468/ab6331}{\emph{Journal of
  Statistical Mechanics: Theory and Experiment} {\bf 2020} (Feb, 2020) 023101}.

\bibitem{javerzat2019fourpoint}
N.~Javerzat, M.~Picco and R.~Santachiara, \emph{{Three- and four-point
  connectivities of two-dimensional critical Q-Potts random clusters on the
  torus}}, \href{http://dx.doi.org/10.1088/1742-5468/ab7c5e}{\emph{Journal of
  Statistical Mechanics: Theory and Experiment} {\bf 2020} (may, 2020) 053106}.

\bibitem{javerzat2020topological}
N.~Javerzat, S.~Grijalva, A.~Rosso and R.~Santachiara, \emph{{Topological
  effects and conformal invariance in long-range correlated random surfaces}},
  \href{http://arxiv.org/abs/2005.11830}{{\tt 2005.11830}}.

\bibitem{Schramm2000}
O.~Schramm, \emph{Scaling limits of loop-erased random walks and uniform
  spanning trees}, \href{http://dx.doi.org/10.1007/BF02803524}{\emph{Israel
  Journal of Mathematics} {\bf 118} (Dec, 2000) 221--288}.

\bibitem{BAUER2002135}
M.~Bauer and D.~Bernard, \emph{{$SLE_{\kappa}$ growth processes and conformal
  field theories}},
  \href{http://dx.doi.org/https://doi.org/10.1016/S0370-2693(02)02423-1}{\emph{Physics
  Letters B} {\bf 543} (2002) 135 -- 138}.

\bibitem{wiegmann2005}
E.~Bettelheim, I.~A. Gruzberg, A.~W.~W. Ludwig and P.~Wiegmann,
  \emph{{Stochastic Loewner Evolution for Conformal Field Theories with Lie
  Group Symmetries}},
  \href{http://dx.doi.org/10.1103/PhysRevLett.95.251601}{\emph{Phys. Rev.
  Lett.} {\bf 95} (Dec, 2005) 251601}.

\bibitem{SAKAI2013429}
K.~Sakai, \emph{{Multiple Schramm–Loewner evolutions for conformal field
  theories with Lie algebra symmetries}},
  \href{http://dx.doi.org/https://doi.org/10.1016/j.nuclphysb.2012.09.019}{\emph{Nuclear
  Physics B} {\bf 867} (2013) 429 -- 447}.

\bibitem{fzpara}
A.~Zamolodchikov and V.~Fateev, \emph{{Nonlocal (parafermion) currents in
  two-dimensional conformal quantum field theory and self-dual critical points
  in $Z_N$-symmetric statistical systems}}, {\emph{JETP} {\bf 62} (August,
  1985) 215}.

\bibitem{Santachiara_2008}
R.~Santachiara, \emph{$\text{SLE}$ in self-dual critical spin systems:
  $\text{CFT}$ predictions},
  \href{http://dx.doi.org/10.1016/j.nuclphysb.2007.09.029}{\emph{Nuclear
  Physics B} {\bf 793} (Apr, 2008) 396–424}.

\bibitem{pisa07}
M.~{Picco} and R.~{Santachiara}, \emph{{Numerical Study on Schramm-Loewner
  Evolution in Nonminimal Conformal Field Theories}},
  \href{http://dx.doi.org/10.1103/PhysRevLett.100.015704}{\emph{Physical Review
  Letters} {\bf 100} (Jan., 2008) 015704},
  [\href{http://arxiv.org/abs/0708.4295}{{\tt 0708.4295}}].

\bibitem{Caselle_2011}
M.~Caselle, S.~Lottini and M.~A. Rajabpour, \emph{{Critical domain walls in the
  Ashkin–Teller model}},
  \href{http://dx.doi.org/10.1088/1742-5468/2011/02/p02039}{\emph{Journal of
  Statistical Mechanics: Theory and Experiment} {\bf 2011} (Feb, 2011) P02039}.

\bibitem{Ikhlef_2010}
Y.~Ikhlef and M.~A. Rajabpour, \emph{{Discrete holomorphic parafermions in the
  Ashkin–Teller model and SLE}},
  \href{http://dx.doi.org/10.1088/1751-8113/44/4/042001}{\emph{Journal of
  Physics A: Mathematical and Theoretical} {\bf 44} (Dec, 2010) 042001}.

\bibitem{fukusumi2017multiple}
Y.~Fukusumi, \emph{{Multiple Schramm-Loewner evolutions for coset
  Wess-Zumino-Witten models}},  \href{http://arxiv.org/abs/1704.06006}{{\tt
  1704.06006}}.

\bibitem{AT43}
J.~Ashkin and E.~Teller, \emph{Statistics of two-dimensional lattices with four
  components}, \href{http://dx.doi.org/10.1103/PhysRev.64.178}{\emph{Phys.
  Rev.} {\bf 64} (Sep, 1943) 178--184}.

\bibitem{N1984}
B.~Nienhuis, \emph{Critical behavior of two-dimensional spin models and charge
  asymmetry in the coulomb gas},
  \href{http://dx.doi.org/10.1007/BF01009437}{\emph{Journal of Statistical
  Physics} {\bf 34} (Mar, 1984) 731--761}.

\bibitem{SaleurAT}
H.~Saleur, \emph{{Partition functions of the two-dimensional Ashkin-Teller
  model on the critical line}}, {\emph{Journal of Physics A: Mathematical and
  General} {\bf 20} (1987) L1127}.

\bibitem{Baxter}
R.~J. Baxter, \emph{{Exactly solved models in statistical mechanics}}.
\newblock Academic Press, London, 1982.

\bibitem{FZ}
V.~A. Fateev and A.~B. Zamolodchikov, \emph{{Conformal quantum field theory
  models in two dimensions having $Z_3$ symmetry}}, {\emph{Phys.~Lett.} {\bf
  92A} (1982) 37}.

\bibitem{Yang-Perk}
H.~{Au-Yang} and J.~H.~H. {Perk}, \emph{{The chiral Potts models revisited}},
  \href{http://dx.doi.org/https://doi.org/10.1007/BF02183338}{\emph{J.~Stat.~Phys.}
  {\bf 78} (1995) 17}.

\bibitem{GC}
A.~Gamsa and J.~L. Cardy, \emph{{Schramm–Loewner evolution in the three-state
  Potts model—a numerical study}},
  \href{http://dx.doi.org/https://doi.org/10.1088/1742-5468/2007/08/P08020}{\emph{Journal
  of Statistical Mechanics: Theory and Experiment} (2007) P08020}.

\bibitem{PS4}
M.~{Picco} and R.~{Santachiara}, \emph{unpublished, 2007}, .

\bibitem{PS3}
M.~{Picco} and R.~{Santachiara}, \emph{{Critical interfaces and duality in the
  Ashkin Teller model}},
  \href{http://dx.doi.org/10.1103/PhysRevE.83.061124}{\emph{Phys.~Rev.} {\bf
  E83} (2011) 061124}, [\href{http://arxiv.org/abs/1011.1159}{{\tt
  1011.1159}}].

\bibitem{Bauer2005}
M.~Bauer, D.~Bernard and K.~Kyt{\"o}l{\"a}, \emph{{Multiple Schramm--Loewner
  Evolutions and Statistical Mechanics Martingales}},
  \href{http://dx.doi.org/10.1007/s10955-005-7002-5}{\emph{Journal of
  Statistical Physics} {\bf 120} (Sep, 2005) 1125--1163}.

\bibitem{bhs17}
V.~Belavin, Y.~Haraoka and R.~Santachiara, \emph{{Rigid Fuchsian systems in
  2-dimensional conformal field theories}},
  \href{http://dx.doi.org/10.1007/s00220-018-3274-x}{\emph{Communications in
  Mathematical Physics} {\bf 365} (2019) 17–60},
  [\href{http://arxiv.org/abs/1711.04361}{{\tt 1711.04361}}].

\bibitem{Blanchard_2013}
T.~Blanchard and M.~Picco, \emph{Frozen into stripes: Fate of the critical
  ising model after a quench},
  \href{http://dx.doi.org/10.1103/physreve.88.032131}{\emph{Physical Review E}
  {\bf 88} (Sep, 2013) 032131}.

\bibitem{WD}
S.~Wiseman and E.~Domany, \emph{{Cluster method for the Ashkin-Teller model}},
  \href{http://dx.doi.org/https://doi.org/10.1103/PhysRevE.48.4080}{\emph{Physical
  Review E} {\bf 48} (1993) 4080}.

\bibitem{Zampara}
A.~B. Zamolodchikov and V.~A. Fateev, \emph{Nonlocal (parafermion) currents in
  two-dimensional conformal quantum field theory and self-dual critical points
  in zn symmetrical statistical models}, {\emph{Zh. Eksp. Teor. Fiz.; (USSR)}
  {\bf 89:2} (Aug, 1985) }.

\bibitem{Dotsenko_2003}
V.~S. Dotsenko, J.~L. Jacobsen and R.~Santachiara, \emph{{Parafermionic theory
  with the symmetry $Z_5$}},
  \href{http://dx.doi.org/10.1016/s0550-3213(03)00066-x}{\emph{Nuclear Physics
  B} {\bf 656} (Apr, 2003) 259–324}.

\bibitem{KNIZHNIK198483}
V.~Knizhnik and A.~Zamolodchikov, \emph{{Current algebra and Wess-Zumino model
  in two dimensions}},
  \href{http://dx.doi.org/https://doi.org/10.1016/0550-3213(84)90374-2}{\emph{Nuclear
  Physics B} {\bf 247} (1984) 83 -- 103}.

\bibitem{DOTSENKO1991547}
V.~Dotsenko, \emph{{Solving the SU(2) conformal field theory using the Wakimoto
  free field representation}},
  \href{http://dx.doi.org/https://doi.org/10.1016/0550-3213(91)90424-V}{\emph{Nuclear
  Physics B} {\bf 358} (1991) 547 -- 570}.

\bibitem{CF_WZW}
P.~Christe and R.~Flume, \emph{{The four-point correlations of all primary
  operators of the $d = 2$ conformally invariant $SU(2) \sigma$-model with
  Wess-Zumino term}},
  \href{http://dx.doi.org/https://doi.org/10.1016/0550-3213(87)90693-6}{\emph{Nuclear
  Physics B} {\bf 282} (1987) 466 -- 494}.

\bibitem{Gori_2018}
G.~Gori and J.~Viti, \emph{Four-point boundary connectivities in critical
  two-dimensional percolation from conformal invariance},
  \href{http://dx.doi.org/10.1007/jhep12(2018)131}{\emph{Journal of High Energy
  Physics} {\bf 2018} (Dec, 2018) 131}.

\end{thebibliography}\endgroup
\bibliographystyle{JHEP}
\end{document}